\documentclass[journal=jacsat,manuscript=communication]{achemso}
\usepackage{graphicx}
\usepackage{color}
\usepackage{titlecaps}
\usepackage{caption}
\usepackage{amssymb}
\usepackage{amsmath}
\usepackage{gensymb}
\usepackage{booktabs}
\usepackage{multicol}
\usepackage[table]{xcolor}
\usepackage{etoolbox}
\makeatletter
\def\acs@contact@details{
E-mail: \acs@email@list
}
\makeatother

\graphicspath{{figures/}}

\title{DL\_POLY Quantum 2.1 software: A suite of real-time path integral methods for the simulation of dynamical properties and vibrational spectra}

\author{Nathan London}
\affiliation[University of Missouri-Kansas City]{Division of Energy, Matter, and Systems, School of Science and Engineering, University of Missouri-Kansas City, Kansas City, MO}
\author{Dil K. Limbu}
\affiliation[University of Missouri-Kansas City]{Division of Energy, Matter, and Systems, School of Science and Engineering, University of Missouri-Kansas City, Kansas City, MO}
\author{Md Omar Faruque}
\affiliation[University of Missouri-Kansas City]{Division of Energy, Matter, and Systems, School of Science and Engineering, University of Missouri-Kansas City, Kansas City, MO}
\author{Farnaz A. Shakib}
\email{shakib@njit.edu}
\affiliation{Department of Chemistry and Environmental Science, New Jersey Institute of Technology, Newark 07102, NJ United States. }
\author{Mohammad R. Momeni}
\email{mmomenitaheri@umkc.edu}
\affiliation[University of Missouri-Kansas City]{Division of Energy, Matter, and Systems, School of Science and Engineering, University of Missouri-Kansas City, Kansas City, MO}


\begin{document}

\begin{abstract}
DL\_POLY Quantum 2.1 is introduced here as a highly modular, sustainable, and scalable general-purpose molecular dynamics (MD) simulation software for large-scale long-time MD simulations of condensed phase and interfacial systems with the essential nuclear quantum effects (NQEs) included. The new release improves upon version 2.0 through the introduction of several emerging real-time path integral (PI) methods, including fast centroid molecular dynamics (f-CMD) and fast quasi-CMD (f-QCMD) methods, as well as our recently introduced hybrid CMD (h-CMD) method for the accurate and efficient simulation of vibrational infrared spectra. Several test cases, including liquid bulk water at 300 K and ice I$_{ \mathrm{h}}$ at 150 K, are used to showcase the performance of different implemented PI methods in simulating the infrared spectra at both ambient conditions and low temperatures where NQEs become more apparent. Additionally, using different salt-in-water (i.e., dilute) and water-in-salt (i.e., concentrated) lithium bis(trifluoromethanesulfonyl)imide (Li-TFSI) aqueous electrolyte solutions, we demonstrate the applicability of our recently introduced h-CMD method implemented in DL\_POLY Quantum 2.1 for the large scale simulation of IR spectra of complex heterogeneous systems. We show that h-CMD can overcome the curvature problem of CMD and the artificial broadening of T-RPMD for the accurate simulation of the vibrational spectra of complex, heterogeneous systems with NQEs included.
\end{abstract}

\section{INTRODUCTION} \label{sec:intro}
\noindent\textbf{INTRODUCTION}
\\\\
Vibrational spectroscopy,\cite{blasiak2017vibrational, roy2013vibrational} specifically infrared (IR) spectroscopy,\cite{paschoal2017vibrational} is immensely important in providing a molecular-level understanding of the structure and dynamics of chemical, material, and biological condensed-phase systems and interfaces.\cite{baiz2020vibrational} Complementary to experimental measurements, atomistic molecular dynamics (MD) simulations can establish structure-property relationships by connecting the vibrational spectral signatures with the complex microscopic structural and dynamical properties at the molecular level.\cite{shepherd2021efficient} To achieve predictive accuracy in simulating vibrational spectra, it is well established that incorporating nuclear quantum effects (NQEs) such as zero-point energy (ZPE) and quantum tunneling is essential.\cite{kapil2024firstprinciples} This becomes particularly important when MD simulations involve light particles and/or lower temperatures, a prominent example being the simulation of bulk water in liquid or ice form or at different interfaces.\cite{marsalek2017quantum} Path integral MD (PIMD) based methods, which perform classical MD in the extended phase space of a ring polymer, are among the most efficient and widely employed approaches for incorporating NQEs into dynamics simulations.\cite{julFeynman2010,janParrinello1984,Hirshberg:2019} Recent advances have shown that approximate real-time PI-based methods can generate spectra that are in excellent agreement with the experimentally measured IR spectra of liquid water and ice over a broad (0-4000 cm$^{-1}$) frequency range in terms of both the positions and shape of the peaks.\cite{kapil2024firstprinciples}  Therefore, by utilizing the predictive nature of PI methods in MD simulations, a wide range of fundamental questions related to deciphering experimental data can be answered with the potential of discovering new phenomena.

Developing a sustainable, secure, and scalable general-purpose software platform that includes both the legacy and emerging real-time PI-based methods, which offer different degrees of accuracy and efficiency, has the potential to advance scientific research across multiple crosscutting physical and material sciences and engineering disciplines. Using the developed software platform, the broader community of both theoretical and experimental research could benefit from these state-of-the-art simulation techniques to investigate the structural and dynamical properties as well as vibrational spectral simulations of different condensed phase and interfacial processes relevant to electrochemical energy conversion and storage technologies. Recently, we have released DL\_POLY Quantum 2.0 \cite{aprLondon2024} as a highly efficient modular software platform for PIMD simulations. Specifically, the DL\_POLY Quantum 2.0 version supports a suite of legacy real-time PI-based methods such as ring polymer MD (RPMD)\cite{Craig:2004,Craig:2005a,julCraig2005}, thermostatted RPMD (T-RPMD),\cite{junRossi2014} centroid MD (CMD)\cite{cmd1,cmd2,cmd3,cmd4} and partially-adiabatic CMD (PA-CMD)\cite{aprHone2006}. While RPMD has been proven successful in calculating various dynamical properties, including reaction rates,~\cite{augCraig2004,Craig:2005a,julCraig2005} it is not suitable for calculating vibrational spectra due to the coupling of the internal modes of the ring polymer to the modes of the physical system, creating spurious peaks in the spectra.~\cite{augHabershon2008} T-RPMD removes these spurious frequencies by coupling the internal modes of the ring polymer to a thermostat without altering their masses. However, depending on the temperature of the simulations and for light nuclei, the calculated IR peaks using T-RPMD are artificially dampened, leading to broad featureless IR spectra, especially in the stretch region. CMD has also been used to calculate vibrational spectra after its introduction,\cite{cmd4,augJang1999,Jang:1999}, but it generates IR peaks that are artificially red-shifted, primarily of stretch bands due to its known ``curvature problem''.~\cite{janIvanov2010}

The development of new real-time PI-based methods is an active research area, with new methods being developed and introduced that extend the accuracy and/or the length- and time-scale of the PI simulations. 
Accordingly, this paper presents a new release of our general-purpose DL\_POLY Quantum software, version 2.1, which has been updated with the emerging state-of-the-art CMD-based methods, including fast CMD (f-CMD),\cite{janHone2005} fast quasi-centroid MD (f-QCMD) \cite{decFletcher2021} as well as our own hybrid CMD (h-CMD)\cite{novLimbu2024} method specifically developed to extend the applicability of the above-mentioned methods to complex heterogeneous systems. 
Introduced by Althorpe and coworkers,\cite{augTrenins2019} QCMD was recently designed to overcome the curvature problem and the artificial red-shifting of vibrational spectra of CMD. This was achieved by defining the centroid that the imaginary time ring polymer is constrained to using an average of curvilinear coordinates instead of Cartesian coordinates.\cite{augTrenins2019}  Although QCMD gives vibrational spectra in remarkably good agreement with quantum mechanical reference calculations, it is prohibitively expensive for systems with more than a few degrees of freedom.\cite{decFletcher2021} Efforts to reduce the computational cost of PIMD-based calculations have been underway for the past few years. Inspired by the ``fast'' version of CMD (f-CMD),\cite{janHone2005}, the fast version of QCMD (f-QCMD)\cite{decFletcher2021} was recently introduced by Manolopoulos and coworkers. Both methods involve using PIMD simulations as a reference to learn a new potential from which classical dynamics can be performed that mimic the CMD and QCMD dynamics, respectively. The f-CMD method learns its potential through force matching to the CMD forces,\cite{janHone2005} while f-QCMD uses iterative Boltzmann inversion (IBI)~\cite{janSoper1996,Reith2003} to learn a correction potential based on reference distribution functions.~\cite{decFletcher2021} DL\_POLY Quantum 2.1 offers these state-of-the-art CMD-based methods in a user-friendly set-up that can be employed by researchers without a need for high levels of coding proficiency. More importantly, DL\_POLY Quantum 2.1 allows large-scale long-time CMD-based simulations corrected for curvature problems in complex condensed phases and interfaces, using our very own hybrid CMD (h-CMD)\cite{novLimbu2024} method, enabling applications of PI methods to real-world dynamical and spectral simulations. The newly introduced h-CMD scheme applies f-QCMD to light and NQE-susceptible particles while using f-CMD for heavier atoms. 

We have structured DL\_POLY Quantum as a highly parallelized, modular, and user-friendly platform, with its latest 2.1 version being the first and only standalone software capable of performing all the above-mentioned flavors of real-time PI methods for systems with thousands of degrees of freedom. This paper is structured as follows. In the next section, we present an overview of the features added/updated in version 2.1 vs. 2.0. The theory details of new methods implemented in the DL\_POLY Quantum 2.1 are briefly provided in section 3. Interested readers are referred to Ref.\citenum{novLimbu2024} for more details and theoretical background. In section 4, we provide a brief tutorial for the users so they can easily utilize DL\_POLY Quantum 2.1 for h-CMD simulations and vibrational spectral analysis. We then demonstrate and discuss the results of three exemplary cases: liquid bulk water at 300 K and ice I$_{ \mathrm{h}}$ at 150 K, as well as aqueous solutions of lithium bis(trifluoromethanesulfonyl)imide (Li-TFSI) at four different concentrations with experimentally measured IR spectra as reference. The conclusions and outlook are provided in the last section.
\begin{figure}[!t]
\centering
\includegraphics[width=\linewidth]{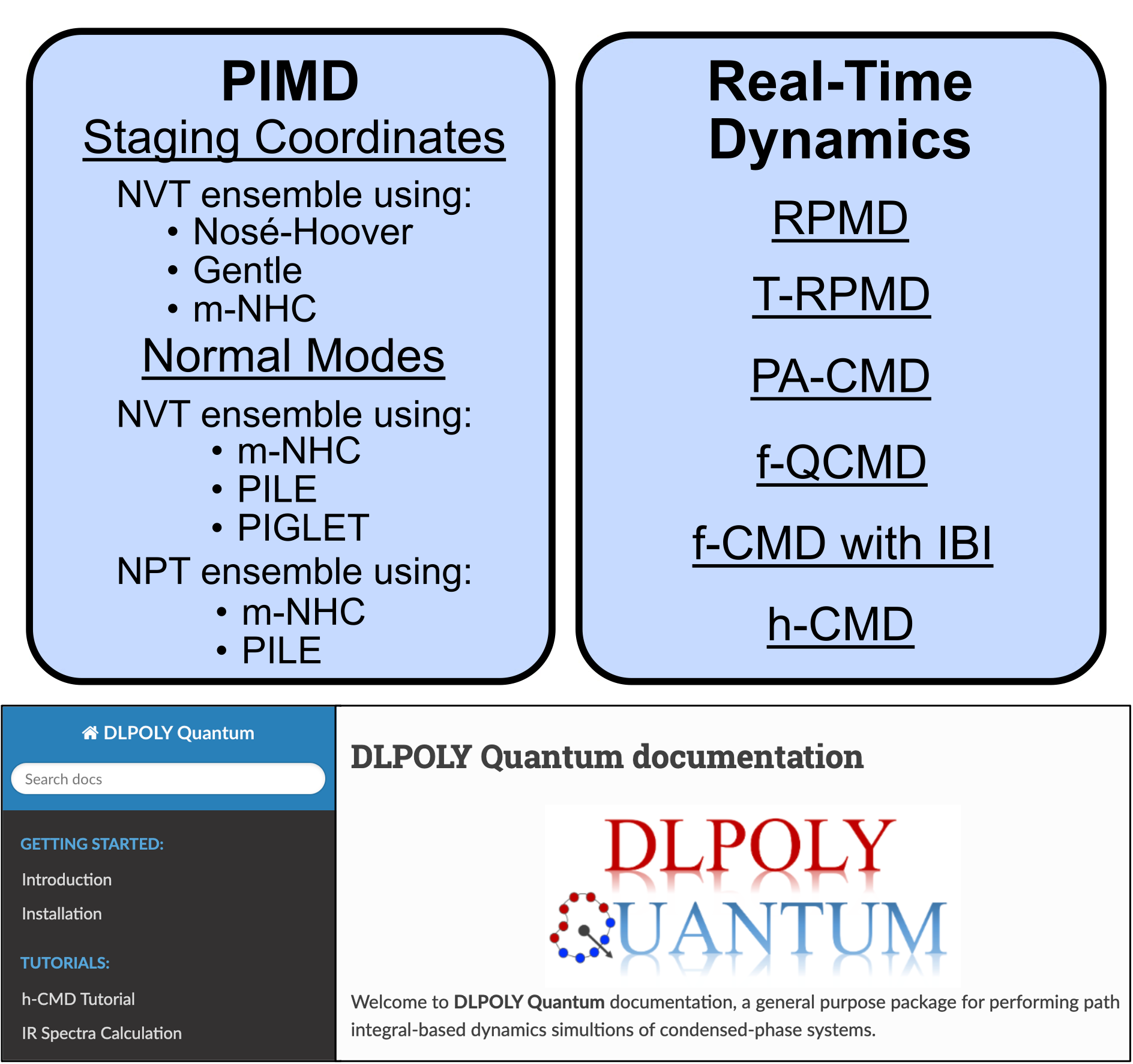}
\caption{Top: List of PIMD functionalities and real-time path integral dynamics methods available in DL\_POLY Quantum software package. Bottom: Snippet of the software documentation website at https://dlpolydocs.readthedocs.io/en/latest/.}
\label{fig:webpage}
\end{figure}
\\\\
\section{PROGRAM FEATURES AND OVERVIEW}
\noindent\textbf{PROGRAM FEATURES AND OVERVIEW}
\\\\
The open-source DL\_POLY Quantum 2.1 software presented in this work is an extensive update to the DL\_POLY Quantum 2.0 package developed and released by some of us recently.\cite{aprLondon2024} This major update contains new functionalities in the form of additional real-time PI-based simulation methods and improving/refining some of the original functionalities. These updates set DL\_POLY Quantum 2.1 apart from other PI codes as it is the only general-purpose software package that includes both legacy PI methods like PIMD, RPMD, PA-CMD, and T-RPMD as well as newly developed and introduced ones, including f-CMD, f-QCMD, and h-CMD. Here, we briefly list the updates featured in version 2.1 of the software.

  \begin{itemize}
    \item The f-QCMD functionality is added through the addition of a new module, \textit{fqcmd\_module.f}. The module adds the ability to generate the reference distribution functions and perform the classical-like simulations under the correction potentials. The f-QCMD simulations are invoked by adding the keyword \textbf{fqcmd} in the CONTROL file. Using this keyword in tandem with a PIMD simulation generates the reference distributions with the use of the quasi-centroids (QCs). Using a classical integration method will apply the correction potentials while calculating the distributions. 
    \item Includes additional Python scripts for calculating the additions to the correction potential during each step in the IBI process. The script is a modified version of the script acquired from the authors of Ref.~\citenum{octLawrence2023}. It is altered to be generalized to systems beyond the ones in Ref.~\citenum{octLawrence2023}. It also directly updates the bond and angle parameters in the DL\_POLY Quantum FIELD file for user convenience. The script generates a new FIELD file with updated parameters and a series of table files to enumerate the inter-molecular correction potentials and forces for each pair type in the system.
    \item Includes a new functionality to read in additional input files associated with f-QCMD simulations. First, the table files for the inter-molecular correction potentials are read in and stored for interpolation during simulations. Second, a unique input file, FQCMD, is introduced that specifies how many unique bond/angle types exist for each molecule in the system and how many times they occur in the molecule. This information is used to properly collect all bonds and angles into the correct intra-molecular distribution functions.
    \item Includes the f-CMD functionality within the new \textit{fqcmd\_module.f}. The method is called using the keyword \textbf{fcmd}. Invoking this option sets an additional boolean that indicates the use of the Cartesian centroid positions when calculating the reference distribution functions. The same table and FQCMD files are also used when performing f-CMD reference distribution functions.
    \item Includes the ability to perform h-CMD simulations. h-CMD simulations are automatically performed when the \textbf{fqcmd} keyword is added to the CONTROL file for systems containing multiple molecules. Note that version 2.1 of DL\_POLY Quantum supports h-CMD simulations, where water molecules are treated with f-QCMD and all other molecules in the systems with f-CMD.
    \item The module used to calculate the values of operators for correlation functions, \textit{correlation\_module.f}, is updated to calculate the total dipole moment of the chosen molecule type instead of molecular dipoles. Additionally, the separate program that calculates the vibrational spectra from the dipole values is updated to calculate the spectra using the dipole derivative instead of the dipole directly. Also, support for using the Hann windowing function~\cite{press_etal:1992} is added.
    \item {Initial momentum distributions for beads in PIMD simulations have been refined. Now, instead of simply resetting the average momentum of beads, the center-of-mass (CoM) of the centroid is specifically recalculated and reset within the \emph{reset\_pimd\_momenta} routine. This adjustment ensures that the system’s CoM momentum is zeroed for each PI simulation, enhancing accuracy in initial momentum distribution and temperature equilibration across beads.}
    \item {The updated PIMD module now computes the CoM momenta of the centroid and removes the mass-weighted contributions from each bead after applying any thermostat attached to the system.~\cite{bussi_2009} This step effectively corrects any drift, preserving CoM consistency across all beads. For CoM calculations, only centroids are used rather than individual beads, ensuring uniform temperature control at the centroid level.}
    \item {The massive Nosé-Hoover chain (m-NHC) thermostat\cite{Tobias:1993,Tuckerman:2010} has been updated to work with a single bead (i.e., classical simulations) using the PIMD module. Previously, the m-NHC thermostat was configured for n-bead PIMD simulations using both staging and normal modes.}
  \end{itemize}

\section{THEORETICAL BACKGROUND} \label{sec:theory}
\noindent\textbf{THEORETICAL BACKGROUND}
\\\\
In this section, we discuss the emerging PI methods that are included within DL\_POLY Quantum 2.1. The discussions of methods carried over from version 2.0 of the software to the current version can be found in Ref.~\citenum{aprLondon2024}. However, we will first give a brief overview of the legacy method of CMD since the new methods have evolved from it.    
\\\\
\subsubsection{Centroid Molecular Dynamics (CMD)}
\textbf{Centroid Molecular Dynamics (CMD)}
\\\\
Introduced by Voth and coworkers,\cite{cmd1,cmd2,cmd3,cmd4} CMD is an approximate method for simulating real-time dynamics using the extended phase space of the ring polymers. In CMD, the dynamics of the particles are performed under the effective mean-field potential of an imaginary-time path-integral whose centroid is constrained to be at the position of the particle. For low dimensional systems, this effective potential can be calculated on a grid prior to dynamical simulations, while for large systems, it must be calculated ``on-the-fly'', which is referred to as adiabatic CMD.~\cite{cmd4} This is achieved by introducing an adiabatic separation between the centroid and internal modes by using the physical mass of the particle as the centroid mass, scaling down the mass of the internal modes of the RP, and attaching the internal modes to a thermostat so that they sample the equilibrium distribution while being constrained to the position of the slower moving centroid.\cite{cmd4} Full adiabatic separation requires a very small mass for the internal modes and, thus, a very small simulation time step for the dynamics to be accurate. This led to the development of the partially adiabatic CMD (PA-CMD) method, where the mass scaling is not as extreme, allowing for a larger time step while still providing accurate dynamics.~\cite{aprHone2006}

The PA-CMD effective Hamiltonian in terms of the free ring polymer normal modes is~\cite{aprHone2006}
\begin{equation}
	H_{\mathrm{PA-CMD}}^0=\sum_{i=1}^{N}\sum_{k=0}^{n-1} 
  \left[ \frac{\tilde{p}_{i,k}^2}{2\sigma_k^2m_n^{(i)}} + \frac{1}{2}m_n^{(i)}\omega_k^2\tilde{q}_{i,k}^2\right],
	\label{eq:PA-CMDHam}
\end{equation}
where $\sigma_k$ is a scaling factor defined as
\begin{equation}
	\sigma_k = 
	\begin{cases}
	1, \quad k=0 \\
	\omega_k/\Omega, \quad k\neq0
	\end{cases}.
	\label{eq:PA-CMDSigma}
\end{equation}
\\\\
\subsubsection{Fast Centroid Molecular Dynamics (f-CMD)}
\textbf{Fast Centroid Molecular Dynamics (f-CMD)}
\\\\
Another way of speeding up CMD simulations is through the use of the f-CMD method.\cite{janHone2005} In f-CMD, a classical-like potential is learned to mimic the effective mean-field potential of CMD through force matching. This new potential can take the form of a standard analytical force field with new parameters,\cite{janHone2005,mayYuan2018} or as a neural network potential (NNP).~\cite{octLoose2022} Once the potential has been learned, it can be used with classical simulations to calculate dynamical properties in line with PA-CMD simulations.

It has been well-documented that vibrational spectra calculated using the above (PA-)CMD and f-CMD methods exhibit red-shifting of peaks, primarily of stretch bands, that is worsened with the lowering of temperatures.\cite{janIvanov2010} This phenomenon, dubbed the ``curvature problem'', is a result of the ring polymer spreading out along the angular coordinates of a system when approaching the inside of stretching coordinates. This stretching reduces the force along the stretching coordinate, lowering the oscillation frequency of that coordinate and red-shifting the spectra. The curvature problem is exacerbated at lower temperatures where a larger number of RP beads is needed and can create an even more spread-out distribution.
\\\\
  \subsubsection{Quasi-Centroid Molecular Dynamics (QCMD)}
\textbf{Quasi-Centroid Molecular Dynamics (QCMD)}
\\\\
To overcome the curvature problem, Althorpe and coworkers introduced the QCMD method.\cite{augTrenins2019} The basic idea behind this method is that the effective mean-field potential is defined along a set of curvilinear coordinates instead of the Cartesian ones. For a given molecule containing $N$ atoms, each being represented with an $n$-bead ring polymer, there is a set of Cartesian coordinates $\{\mathbf{q}_1,\dots,\mathbf{q}_N\}$, where $ \mathbf{q}_i = \{q_{i1},\dots,q_{in}\} $ represents the coordinates of all the beads of the ring polymer (atom) $i$. Additionally, the set of $L$ bonds and $M$ angles of the molecule can be defined in terms of the bond lengths and angle values of each replica of the system, $ \{r_{1},\dots,r_{L}\} $ and $\{\theta_{1},\dots,\theta_{M}\}$. A set of curvilinear coordinates, $ \{R_{1},\dots,R_{L}\} $ and $\{\Theta_{1},\dots,\Theta_{M}\}$, can be found by bead-averaging as,~\cite{augTrenins2019}
  \begin{equation}
    R_l = \frac{1}{n}\sum^{n}_{\alpha=1} r_{l,\alpha},
    \label{eq:radCon}
  \end{equation}
  \begin{equation}
    \Theta_{m} = \frac{1}{n} \sum^{n}_{\alpha=1} \theta_{m,\alpha}.
    \label{eq:thetaCon}
  \end{equation}

  For practical use in simulations, it is necessary to convert these curvilinear coordinates back into Cartesian
  coordinates for each atom. Because the relationship between the Cartesian and polar bead coordinates is non-linear,
  this new set of coordinates called the quasi-centroids or QCs, $
  \{\overline{\mathbf{Q}}_1,\dots,\overline{\mathbf{Q}}_N\} $
      does not equal that of the Cartesian centroids,\cite{augTrenins2019}
      $\left\{ 
        {\mathbf{Q}}_1,\dots
        {\mathbf{Q}}_N
      \right\}$,
      defined as
      \begin{equation}
        \mathbf{Q}_\mathrm{X} = \frac{1}{n}\sum^{n}_{\alpha=1}\mathbf{q}_\mathrm{X}.
      \end{equation}
Nevertheless, as the ring polymer distribution becomes more compact, the two sets become
      closer in value and are exactly equal in the classical limit.\cite{augTrenins2019}

For condensed phase systems, the relative positions of the atoms in different molecules are required. The curvilinear coordinates used to define the QCs are enough to define the locations of atoms in a molecule but do not position or orient the molecule in 3D space. Using an Eckart-like frame rotation\cite{aprEckart1935,marWilson1980} and the Cartesian centroids as the ``reference'' frame, condensed phase systems can be simulated.\cite{augTrenins2019} The original QCMD method uses an adiabatic implementation to simulate condensed phase systems, where two systems are evolved simultaneously.~\cite{augTrenins2019} The first is the system of interest, and the second is the ring polymer\cite{marRyckaert1977,octAndersen1983,Tuckerman:2010} constrained so that its QCs match the positions of the corresponding atoms of the first system and generates the effective mean-field potential that the first system is evolved under. The forces between atoms of different molecules are introduced through rotational forces, with approximations to the torque being needed when calculating the QC forces.\cite{augTrenins2019,novTrenins2022}
\\\\
  \subsubsection{Fast Quasi-Centroid Molecular Dynamics (f-QCMD)}
\textbf{Fast Quasi-Centroid Molecular Dynamics (f-QCMD)}
\\\\
While successful in simulating accurate spectra with NQEs and avoiding the curvature problem,~\cite{augTrenins2019} the adiabatic QCMD implementation requires a small simulation time step, and the use of rotational forces can be difficult to generalize. To make the method more feasible, Manolopoulos and coworkers introduced f-QCMD.\cite{decFletcher2021} Inspired by f-CMD, f-QCMD trains a corrective potential from which classical simulations can be performed that mimics adiabatic QCMD dynamics. The potential used in f-QCMD has the form\cite{decFletcher2021,octLawrence2023}
  \begin{equation}
    \label{eq:qc-pot}
    V_{\mathrm{qc}}(\mathbf{r}) = V_{\mathrm{cl}}(\mathbf{r}) + \Delta V_{\mathrm{intra}}(\mathbf{r}) +
      \Delta V_{\mathrm{inter}}(\mathbf{r}),
  \end{equation}
  where $V_{\mathrm{cl}}(\mathbf{r})$ is the standard classical potential of the system, and $\Delta V_{\mathrm{intra}}(\mathbf{r})$
  and $\Delta V_{\mathrm{inter}}(\mathbf{r})$ are the correction terms for intra- and inter-molecular interactions,
  respectively. The intra-molecular correction, for a system of $K$ molecules, is approximated as 
  corrections in terms of curvilinear
  coordinates,\cite{decFletcher2021,octLawrence2023}
  \begin{equation}
    \Delta V_{\mathrm{intra}}(\mathbf{r}) \simeq \sum^{K}_{I=1} \sum^{L}_{l=1} \Delta V_R(R_{Ii}) +
    \sum^{K}_{I=1}\sum^{M}_{m=1} \Delta V_\Theta(\Theta_{Ii}),
    \label{eq:intra-correct}
  \end{equation}
  and the inter-molecular correction is taken as a sum of pairwise contributions,\cite{decFletcher2021,octLawrence2023}
  \begin{equation}
    \Delta V_{\mathrm{inter}}(\mathbf{r}) \simeq \sum^{K}_{I=1} \sum^{K}_{J\neq I} \sum_{\mathrm{X} \in I} 
    \sum_{\mathrm{Y}\in J}
    \Delta V_\mathrm{XY}(|\overline{\mathbf{Q}}_{\mathrm{X}I} - \overline{\mathbf{Q}}_{\mathrm{Y}J}|),
    \label{eq:inter-correct}
  \end{equation}
  where the sums over $I$ and $J$ are over molecules, with $J\neq I$, and last two sums are over the atoms
  of type X and Y in molecules $I$ and $J$, respectively.

  Instead of using force matching to obtain the correction potentials, f-QCMD uses iterative Boltzmann inversion 
  (IBI)~\cite{janSoper1996,Reith2003} with a set of two types of distribution
  functions.~\cite{decFletcher2021,octLawrence2023} The first type of distribution function is radial distribution functions (RDFs),
  \begin{equation}
    g_{\mathrm{XY}}(r) = \frac{V(1+\delta_{\mathrm{XY}})}{4\pi r^2 N_{\mathrm{X}}N_{\mathrm{Y}}} 
    \sum_{I=1}^{N_{\mathrm{X}}} \sum_{J=1}^{N_{\mathrm{Y}}}{\vphantom{\sum}}'
      \left\langle
      \delta(r-|\overline{\mathbf{Q}}_{\mathrm{X}I}-\overline{\mathbf{Q}}_{\mathrm{Y}J}|)\right\rangle 
  \end{equation}
  where $V$ is the volume of the simulation cell, $N_{\mathrm{X}}$ and $N_{\mathrm{Y}}$ are the number of atoms of type X and Y, respectively, $\delta_{\mathrm{XY}}$ equals 1 if X and Y are the same atom type, and the prime on the
  second summation indicates that atoms $I$ and $J$ are within different molecules. 

  f-QCMD enforces the Eckart-like conditions through the use of a rotation matrix,\cite{octLawrence2023}, which 
  is chosen to minimize the sum of the mass-weighted squared deviations,
  \cite{augTrenins2019,novTrenins2022,octLawrence2023}
  \begin{equation}
    w(\mathbf{U}) = \sum_j m_j 
    \left[ \left(\mathbf{Q}_j - \mathbf{Q}_{\mathrm{CoM}}\right)
      - \mathbf{U}\left(\overline{\mathbf{Q}}_j - \overline{\mathbf{Q}}_{\mathrm{CoM}}\right)
    \right]^2,
  \end{equation}
  where the sum runs over the atoms in the molecule and $\mathbf{Q}_{\mathrm{CoM}}$ and 
  $\overline{\mathbf{Q}}_{\mathrm{CoM}}$ are the CoMs of the molecule using the Cartesian centroid and
  QC coordinates, respectively. The transformed Cartesian QC coordinates are then found using
  \begin{equation}
    \overline{\mathbf{Q}}^*_j = 
    \mathbf{Q}_{\mathrm{CoM}}
      + \mathbf{U}\left(\overline{\mathbf{Q}}_j - \overline{\mathbf{Q}}_{\mathrm{CoM}}\right).
  \end{equation}
  The rotated QC coordinates are positioned such that the molecule's CoM is the same for both the centroid and QC atom positions and oriented to minimize the difference between the centroid and QC atom positions while constrained to the curvilinear coordinates. The rotation matrix $\mathbf{U}$ is easily found
  using quaternion algebra.\cite{aprKrasnoshchekov2014,octLawrence2023}

  The second type of distribution functions used for IBI are the intra-molecular
  distributions,\cite{decFletcher2021,octLawrence2023}
  \begin{equation}
    \rho_{R_{l}}(r) = \frac{1}{K}\sum_{I=1}^K\left\langle \delta(r-R_{Il}) \right\rangle
  \end{equation}
  and
  \begin{equation}
    \rho_{\Theta_{m}}(\theta) = \frac{1}{N}\sum_{I=1}^K\left\langle \delta(\theta-\Theta_{Im}) \right\rangle,
  \end{equation}
  where $R_{Il}$ and $\Theta_{Im}$ are the values from Eqns.~\ref{eq:radCon} and~\ref{eq:thetaCon} for molecule
  $I$, respectively. For molecules with multiples of the same bond or angle (e.g., the two OH bonds in water), all similar bonds and angles are combined into the same distribution functions.

  The IBI process starts with PIMD simulations used to generate reference values for the distributions described above.
  From there, a classical dynamics simulation is performed, taken as an iteration zero, with the same distribution
  functions calculated. For the $i$-th iteration of the process, the correction potential is updated using the previous
  iteration's distribution functions,
  \begin{equation}
    \begin{split}
      \Delta V_R^{(i+1)}(r) &= \Delta V_R^{(i)}(r) - \frac{1}{\beta}\mathrm{ln}\left( 
        \frac{\rho_R^\mathrm{exact}(r)}{\rho_R^{(i)}(r)}\right)\\
        \Delta V_\Theta^{(i+1)}(\theta) &= \Delta V_\Theta^{(i)}(\theta) - \frac{1}{\beta}\mathrm{ln}\left( 
        \frac{\rho_\Theta^\mathrm{exact}(\theta)}{\rho_\Theta^{(i)}(\theta)}\right)\\
          \Delta V_\mathrm{XY}^{(i+1)}(r) &= \Delta V_\mathrm{XY}^{(i)}(r) - \frac{1}{\beta}\mathrm{ln}\left( 
          \frac{g_\mathrm{XY}^\mathrm{exact}(r)}{g_\mathrm{XY}^{(i)}(r)}\right),
    \end{split}
    \label{eq:ibi-standard}
  \end{equation}
where the exact distributions are taken to be those from the reference PIMD simulations. The QCs in the PIMD simulations are only calculated for use in obtaining the distribution functions and have no bearing on the dynamics. The distribution functions are found simply through binning histograms and are not involved in the dynamics. For the following classical simulations of the IBI process, the RDFs are calculated using force sampling\cite{octRotenberg2020} to allow for shorter simulations.

  For simplicity, the intra-molecular correction potential is taken to have the same functional form as the classical potential for the system with updated parameters.\cite{decFletcher2021,octLawrence2023} The inter-molecular corrections are treated on a grid of $r$ values that are interpolated during the simulation and added to the standard non-bonded interactions. Due to the statistical noise for small values within the RDFs, the IBI process can run into convergence issues.\cite{octLawrence2023} As such, the regularized form of IBI is used for the inter-molecular corrections. The updates to the corrections potentials are now
  \begin{equation}
    \Delta V^{(i+1)}_{\mathrm{XY}}(r) = \Delta V^{(i)}_{\mathrm{XY}}(r) - \frac{1}{\beta}
    \ln\left(\frac{g_{\mathrm{XY}}^{\mathrm{exact}}(r)+\varepsilon G_{\mathrm{XY}}}
    {g_{\mathrm{XY}}^{(i)}(r) + \varepsilon G_{\mathrm{XY}}}\right),
    \label{eq:ibi-reg}
  \end{equation}
  where $\varepsilon$ is a positive scalar parameter and $G_\mathrm{XY}$ is the maximum value of the two distribution functions, allowing for the same value of $\varepsilon$ for all RDFs. Increasing values of $\varepsilon$ reduces the magnitude of the correction potential, helping to stabilize the IBI process.~\cite{octLawrence2023} While the IBI process does need to be performed for a given system at a given temperature, once the correction potentials for it are learned, it can be used for any dynamical simulation of that system.
\\\\
  \subsubsection{Hybrid Centroid Molecular Dynamics (h-CMD)}
\textbf{Hybrid Centroid Molecular Dynamics (h-CMD)}
\\\\
  While f-QCMD is able to create vibrational spectra with similar accuracy to AQCMD simulations, it still has difficulty when applied to complex systems, particularly those containing large complex molecules or materials. This comes from applying the Eckart-like conditions using a rotation matrix, as it requires an initial set of QC coordinates. For large molecules, going from the curvilinear coordinates to QCs can be challenging to generalize.

As a way to generalize f-QCMD to more complex systems, some of us have very recently introduced the hybrid CMD (h-CMD) method.\cite{novLimbu2024} The basis of this method is that for complex condensed phase and interface systems, often only a limited number of degrees of freedom suffer greatly from the curvature problem and need to be treated at the QCMD level, while the rest can be treated with CMD. To test this hypothesis, using an interfacial system of D$_2$O molecules confined in the ZIF-90 metal-organic framework (MOF), we have shown that a good agreement with experimental vibrational spectra can be obtained if the water and the framework are treated with the f-QCMD and f-CMD methods, respectively, within the h-CMD scheme.\cite{novLimbu2024} More specifically, we showed that not only the position of the O-D stretch peaks but their doublet shape characteristic of the interfacial water is reproduced using the h-CMD method. This is in contrast to the red-shifted peaks in PA-CMD due to the curvature problem or the loss of the doublet feature due to the artificial broadening of the spectra as obtained from the T-RPMD method.\cite{novLimbu2024}
We would like to emphasize here again that the h-CMD method offers a path forward for cases where treating the entire system with the f-QCMD method is not feasible due to difficulties in defining curvilinear coordinates. To bring f-CMD on equal footing with f-QCMD, the f-CMD implementation in h-CMD uses the same form of correction potentials as f-QCMD and is found with IBI instead of force matching. The only difference between the f-QCMD and f-CMD calculations within the h-CMD scheme is that the reference distribution functions for the f-CMD molecules are found using the Cartesian centroids instead of the QCs.

As demonstrated with D$_2$O in the ZIF-90 MOF system, h-CMD is able to produce accurate vibrational spectra for complex interfacial systems.~\cite{novLimbu2024} Here, we showcase h-CMD in simulating the vibrational spectra of four different concentrations of Li-TFSI in aqueous solutions compared to the available experiment. Detailed step-by-step instructions are provided for one of the Li-TFSI concentrations as an example. 
\\\\
\subsection{Infrared Spectra Calculations}
\textbf{Infrared Spectra Calculations}
\\\\
The IR spectra calculations in version 2.1 have been updated to use the total cell dipole-derivative as, 
\begin{equation}
    n(\omega)\alpha(\omega) = \frac{\pi\beta}{3cV\epsilon_0}\tilde{I}(\omega),
\end{equation}
where $n(\omega)$ is the refractive index, $\alpha(\omega)$ is the absorption cross-section, and $\tilde{I}(\omega)$ 
is the Fourier transform of the total cell dipole-derivative autocorrelation function,
\begin{equation}
\tilde{I}(\omega) = \frac{1}{2\pi}\int_{-\infty}^{\infty}dt\ \mathrm{e}^{-i\omega t} \langle
       \dot{\boldsymbol{\mu}}(0)\cdot\dot{\boldsymbol{\mu}}(t) \rangle f(t)
\end{equation}
where $\langle ... \rangle$ denote the microcanonical ensemble average, $\dot{\boldsymbol{\mu}}$ is the time
derivative of the total dipole moment of the system, and $f(t)$ is a Hann window function~\cite{press_etal:1992}
\begin{equation}
  f(t) = 
  \begin{cases}
    \cos^2\left( \frac{\pi t}{2\tau}\right)\quad &|t| \leq \tau \\
      0 & |t| > \tau
  \end{cases}  
\end{equation}
 where $\tau$ is a cut-off time chosen as appropriate for the correlation function of interest.
\\\\
\section{h-CMD Tutorial}
\noindent\textbf{h-CMD TUTORIAL}
\\\\
\\\\
\textbf{Setup and PIMD reference calculations}
\\\\ 

\begin{figure}[!t]
    \centering
    \includegraphics[width=\linewidth]{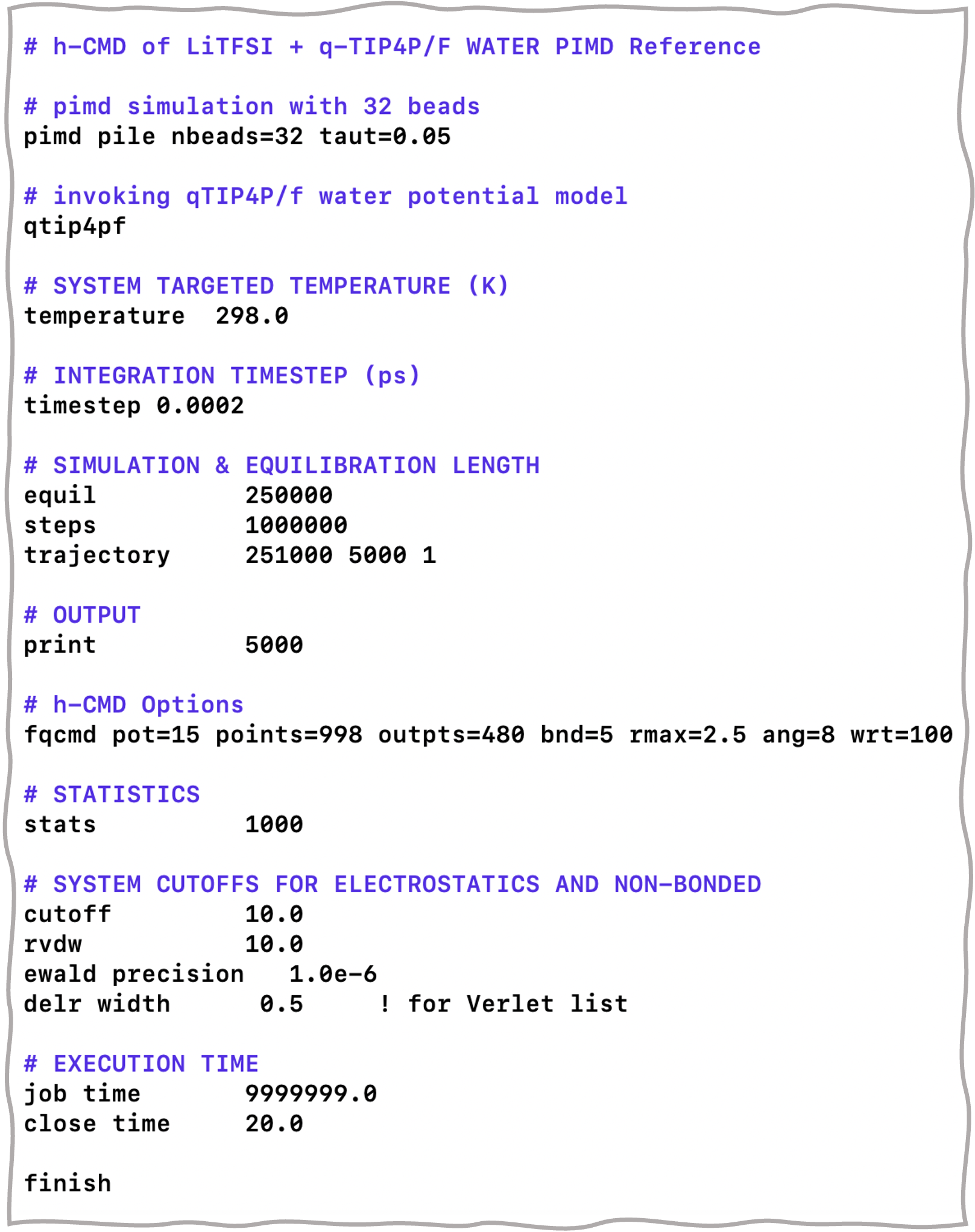}   
    \caption{Example CONTROL file for performing PIMD reference simulations for the h-CMD method.}
    \label{fig:pimd-control}
\end{figure}

To perform h-CMD simulations, additional keywords are added to the CONTROL file, which can be seen in
  Figure~\ref{fig:pimd-control}. This CONTROL file is for performing the initial PIMD reference calculations, so the
  keyword \textbf{pimd} is used to indicate the integrator to be used. 

  The specification that this is part of an h-CMD simulation is through the \textbf{fqcmd} keyword. DL\_POLY Quantum
  will automatically treat the simulation as h-CMD when this keyword is added for systems with multiple molecule types
  defined in the FIELD file. Version 2.1 will treat the final molecule type, assumed to be water, at the f-QCMD level
  and any other molecule types at the f-CMD level.

  The options following the \textbf{fqcmd} keyword specify the details of how the h-CMD simulation will work. The first
  option, \textbf{pot}, indicates the number of non-bonded pair interactions that will employ a correction potential.
  Each pair will have an associated TABLE file, named PAIR\_$<$i$>$\_POT.TABLE. The first line of the table file
  contains the atom names of the pair as defined in the FIELD file. The rest of the file contains a grid of $r$ values
  indicating the distance between the pair of atoms and the associated correction potential and force at that distance.
  For the PIMD reference simulations, the potential and force values are ignored as the dynamics are the same as
  standard PIMD simulations, but the pair names are read from the files to determine which pairs to calculate RDFs for.
  The \textbf{points} option specifies the number of grid points that are in the TABLE files.

  h-CMD simulations generate 3 additional output files: AVERAGE\_INTRA.D, AVERAGE\_ANGLE.D, and AVERAGE\_RDF.D. The first two files include the intra-molecular distributions for the bonds and angles of the system, respectively, and the last
  file is the RDFs. The number of points in these output files is indicated by the \textbf{outpts} option.

\begin{figure}[t]
    \centering
    \includegraphics[width=0.25\linewidth]{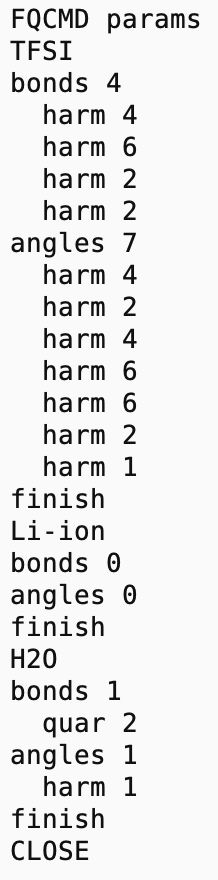}   
    \caption{Example FQCMD file for performing h-CMD simulations for the Li-TFSI system.}
    \label{fig:fqcmd}
\end{figure}
  One more h-CMD-specific input file is needed, the FQCMD file. The FQCMD file for Li-TFSI is shown in Figure~\ref{fig:fqcmd}. In this file, the intra-molecular details needed for h-CMD for each molecule type are specified. The molecules must be in the same order as in the FIELD file. For each molecule, information on the bonds and angles is outlined. The number after the \textbf{bonds} option indicates the number of unique bond types in the molecule. Each following line gives information about the type of bond (harmonic or quartic in this system) and the number of times that bond type occurs. For the water molecules, there is one bond type that appears twice in each molecule, the OH bond. The same idea applies to the \textbf{angles} option, specifying the number of unique angle types in the molecule and how many times each occurs. The information in the FQCMD file is used to ensure that all bonds and angles in the molecules that share the same force field parameters are included in the correct distribution functions.
  
  The total number of unique bonds in the system is included in the CONTROL file with the \textbf{bnd} option, with the \textbf{rmax} option specifying the maximum value $r$ value for the bond distribution function. The maximum value for the angle distribution function is set to be $\pi$, and the maximum for the RDFs is the non-bonded cut-off value in the CONTROL FILE. The number of unique angles is given by the \textbf{ang} option. The final option under the \textbf{fqcmd} keyword is the \textbf{wrt} option, which specifies how many timesteps there should be between calculations for the distribution functions. Collection of the distribution function values does not start until after equilibration of the system as defined by the number of steps following \textbf{equil} keyword.

  It is often advantageous to perform multiple PIMD reference simulations starting from different initial configurations to allow for better distribution function statistics with shorter simulation times. Once all the simulations are finished, the distribution function files can be averaged together using the provided utility program.

\textbf{Iteration 0}

\begin{figure}[!t]
    \centering
    \includegraphics[width=\linewidth]{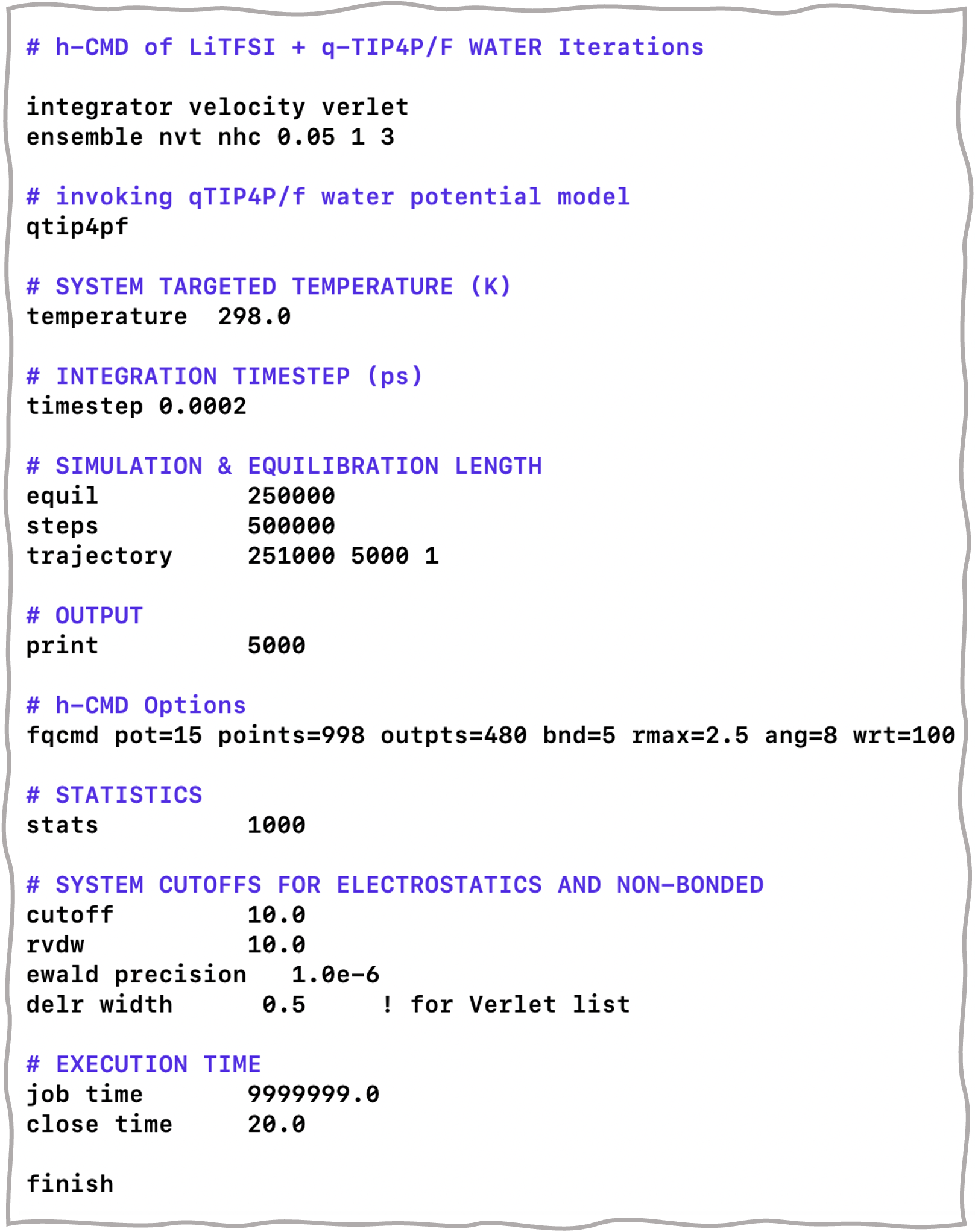}   
    \caption{Example CONTROL file for performing the classical-like simulations during IBI process.}
    \label{fig:iter-control}
\end{figure}

To begin the IBI process, iteration zero is needed where the dynamics are performed classically using the base force field potential. This is equivalent to performing h-CMD with the correction potentials set to zero. Most of the setup work done for the PIMD reference simulations are able to be reused here. The FQCMD file requires no changes. The TABLE files need to have all values of the potential and force set to zero so that the dynamics are not altered from the base potential. The CONTROL file remains very similar to the previously used, as shown in Figure~\ref{fig:iter-control}. The integrator is changed to invoke the velocity Verlet algorithm~\cite{julVerlet1967}, and the NVT ensemble is maintained through a m-NHC thermostat. Additionally, the simulation length is shortened as the RDFs are calculated using force sampling when classical dynamics are used.

\textbf{IBI iterations}

\begin{figure}[!t]
    \centering
    \includegraphics[width=0.4\linewidth]{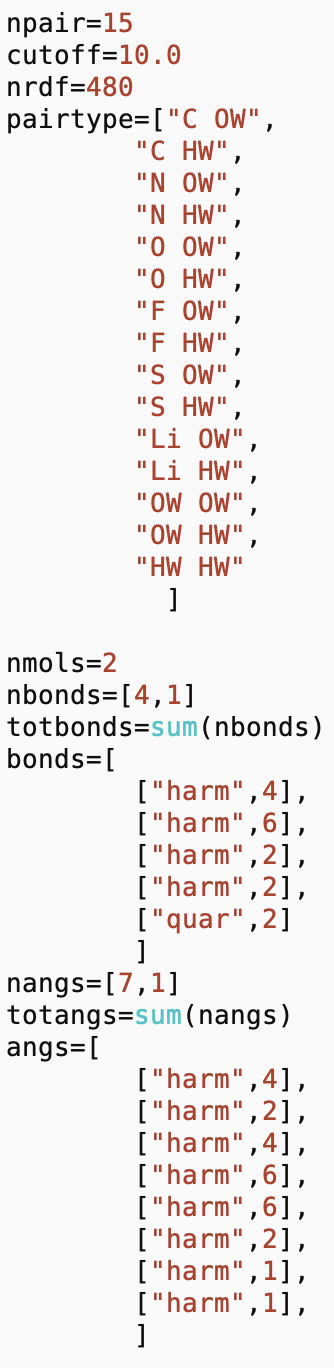}   
    \caption{Snippet of the IBI code detailing the non-bonded pairs and intra-molecular details for the Li-TFSI system.}
    \label{fig:ibi}
\end{figure}

After the PIMD reference and iteration 0 simulations are done, the IBI process can begin. Between each IBI iteration, the correction potentials are updated. This update is done using a separate Python script, which is an edited version of the code provided by the authors of Ref.~\citenum{octLawrence2023} to generalize to systems beyond pure water. A small portion of the script, as depicted in Figure~\ref{fig:ibi} must be edited to detail the system of interest. Most of the details parallel the FQCMD and CONTROL files, with some additional information required. Of particular note are the different arrays in the script. The \textbf{pairtype} array holds the list of non-bonded pairs for which the correction potential is calculated. This list should be in the same order as TABLE files and will be used to write the header line for the output. The \textbf{nbonds} array details the number of unique bonds in each of the molecule types in the system. Like in the FQCMD file, the \textbf{bonds} and \textbf{angles} arrays store information about the type of each unique bond and its occurrence. In these arrays, however, all molecules are combined into one, but they should be in the same order as the FQCMD file.

When running the IBI script, two additional files are required. the first is the FIELD file used for iteration zero, renamed as OLD\_FIELD, from which the bond and angle force field parameters will be read. The second is the PREVIOUS\_POTENTIAL.DAT file, which combines the correction potentials for all the pairs used in the previous iteration into one file. For iteration 0, these will all be zero, but the file is still needed to be run.

After the IBI script is run, a series of new files is created. First is the NEW\_FIELD file, which contains the updated intra-molecular force field parameters as defined by Eq.~\ref{eq:ibi-standard}. A NEW\_POTENTIAL.DAT file is created with the updated inter-molecular correction potentials as defined by Eq.~\ref{eq:ibi-reg}. Additionally, a new series of TABLE files is generated containing the new potentials for each pair type in a separate file, as well as the associated forces.

These new files are then used to run the simulations for the next iteration. Starting with the true iteration one, the dynamics are now classical dynamics under the correction potentials working to mimic PI-base dynamics. The process of running the simulations and fitting the updates to the potentials is repeated until the distribution functions sufficiently match those of the PIMD reference simulations. To simplify the back-and-forth nature of this process, a set of scripts is provided with the code to automatically set up and run IBI iterations. These scripts are designed to work on high-performance computing clusters with SLURM schedulers, with the jobs for later iterations requiring the previous ones to finish before running.

\textbf{Production Simulations}

Once the correction potentials are converged, they can be used for any future simulations of that system. The CONTROL files for these simulations should still contain the same options under the \textbf{fqcmd} keyword as previous simulations, but the value \textbf{wrt} option can be increased to reduce computation time as the distribution functions are no longer needed but are still being calculated.

We have developed a comprehensive documentation website that provides detailed tutorials on PI methods implemented in DL\_POLY Quantum 2.1. Figure~\ref{fig:webpage} displays a screenshot of the website. For further details, please visit the documentation at https://dlpolydocs.readthedocs.io/.
\\~\\
\section{Simulation Details}\label{sec:sim_details}
\textbf{SIMULATION DETAILS}
\\\\
\subsubsection{Bulk Water and Ice I$_{ \mathrm{h}}$}\label{sec:ir_details}
\textbf{Bulk Water and Ice I$_{ \mathrm{h}}$}
\\\\
All simulations for bulk water and ice I$_{ \mathrm{h}}$ presented in this work use the q-TIP4P/F water potential~\cite{julHabershon2009}, which is a flexible 4-site water model designed for use in path-integral simulations. This potential is the base classical potential, $V_\mathrm{cl}(\mathbf{r})$, for f-QCMD and f-CMD methods. Liquid bulk water simulations are performed at 300 K with 216 water molecules in a cubic box with periodic boundary conditions. The box length is 18.648 \AA\ to achieve a density of 0.997 g/cm$^3$, and the cut-off radius for non-bonded interactions is 9 \AA. Simulations are performed with a 0.2 fs time step.

The details for f-QCMD and f-CMD simulations are identical, with the only difference being in how the reference distribution functions are calculated. The PIMD reference calculations are performed using $n=32$ beads. The NVT ensemble is achieved using the PILE thermostat~\cite{sepCeriotti2010}, with a time constant of 0.05 ps. To ensure the distribution functions are properly converged, 50 independent NVT simulations are performed and averaged over. Each simulation involves 50 ps of equilibration and then 150 ps of data collection, with the distribution functions calculated every 100 steps.

During the IBI process, the distribution functions are generated from classical-like simulations under the correction potentials. Like with the reference calculations, a set of 50 independent trajectories are averaged over, but by using force sampling for the RDFs, shorter simulations involving 50 ps of equilibration and 50 ps of data gathering are allowed. The regularization parameter, $\varepsilon$, is set to 1, and a total of 30 iterations are performed to converge the correction potentials. For these simulations, the NVT ensemble is obtained using the m-NHC thermostat with a chain length of 3 and a time constant of 0.05 ps.
As with previous work,\cite{novLimbu2024}, the fitting of the correction potential is done through a modified version of the code provided by the authors of Ref.~\citenum{octLawrence2023}. 

Spectra calculations for all methods follow the same basic procedure. Configurations for dynamics trajectories are generated from NVT simulations at 300 K that involve 50 ps of equilibration and then 50 ps of sampling, with configurations taken every 1 ps. Sampling for classical, f-QCMD, and f-CMD are all done using a m-NHC thermostat with a chain length of 3 and a time constant of 0.05 ps. Sampling for other PI methods is done with PIMD simulations using 32 beads and the PILE thermostat with a time constant of 0.05 ps. Real-time dynamics for all methods are done with 50 independent NVE trajectories for 20 ps. RPMD simulations are performed with 32 beads with no thermostatting. PA-CMD simulations use 32 beads, and the internal RP modes are thermostatted at 300 K with m-NHC thermostats of length 3 and time constants of 0.05 ps. T-RPMD simulations are done with 32 beads and the internal modes are thermostatted with the PILE thermostat.

IR spectra for the liquid bulk water are calculated using the total dipole derivative correlation function. The first 10 ps of each trajectory is ignored, with the second 10 ps being used. A Hann window with $\tau=0.6$ ps is applied to the correlation function. The simulations for ice at 150 K are very similar, with a few key differences. The simulation cell contains 96 water molecules arranged to form a single I$_\mathrm{h}$ crystal in an orthorhombic box with side lengths of 15.65 \AA, 13.56 \AA, and 14.76 \AA, and periodic boundary conditions are applied with a non-bonded interaction cut-off radius of 6.75\AA. All simulations use a time step of 0.2 fs apart from the PA-CMD simulations, which use a 0.1 fs time step. All PI simulations, including the reference simulations for f-QCMD and f-CMD, use 64 beads. The regularized IBI steps use a regularization parameter of 5 to ease convergence. Finally, the Hann window uses a cut-off value of 0.5 ps.
\\\\
\textbf{Li-TFSI Aqueous Electrolyte Solutions}
\\\\
Aqueous Li-TFSI solutions with four different concentrations of 1m, 5m, 10m, and 20m were prepared by randomly placing the appropriate number of Li-TFSI and water molecules in a cubic box of appropriate side lengths using PACKMOL (see Table~\ref{tab:density})\cite{packmol:Martnez:2009}. All systems underwent simulated annealing at 1000 K, 500 K, and 298 K for 50 ps, 30 ps, and 20 ps, respectively, to ensure a sufficiently randomized initial structure. This was followed by a 2 ns equilibration at 298 K in the isothermal-isobaric (NPT) ensemble to determine the system density. The simulated densities, provided in Table~\ref{tab:density}, showed good agreement with experimental values calculated from concentrations reported in Ref.~\citenum{litfsi_2017_density} and AIMD simulated result of Ref.~\citenum{litfsi_spectra_aimd}.
\begin{table}[!h]
\caption{Compositions and simulated densities of aqueous Li-TFSI solutions.}
\centering
\setlength{\tabcolsep}{2.5mm}
\begin{tabular}{c c c c c}
\hline
Molality & Number of & Number of & Density \\
(m) & Li-TFSI & Water & (g/cm$^3$) \\
\hline 
1 & 5 & 275 & 1.149 \\
5 & 18 & 200 & 1.459 \\
10 & 27 & 150 & 1.593 \\
20 & 36 & 100 & 1.715 \\
\hline
\end{tabular}    
\label{tab:density}
\end{table}
This was followed by NVT sampling, which was used for NVE simulations and IR spectra. The simulation protocols for f-CMD and h-CMD were identical to those used for bulk water, the main difference being the number of trajectories utilized in the reference PIMD and real-time NVE dynamic simulations. For the Li-TFSI systems, the reference distribution functions were computed from 20 independent PIMD runs. During IBI fitting, the regularization parameter $\varepsilon$ was set to 5, with 30 iterations performed to ensure convergence of the correction potentials. Real-time dynamics for all methods were conducted with 20 independent NVE trajectories, each run for 50 ps of simulation time. The IR spectra were computed using the total dipole derivative correlation function. The initial 5 ps of each trajectory were discarded to ensure the system reaches equilibrium, particularly for methods involving thermostatting, and to eliminate any artifacts arising from transient dynamics. The subsequent 45 ps of the trajectory were used for analysis. A Hann window with $\tau$ = 0.6 ps was applied.
\\\\
  \section{Results and Discussion} \label{sec:results}
\textbf{RESULTS and DISCUSSION}
\subsection{Bulk Water and Ice I$_{ \mathrm{h}}$}
\\\\
\textbf{Bulk Water and Ice I$_{ \mathrm{h}}$}
\\\\
\begin{figure}[!t]
    \centering
    \includegraphics[width=\linewidth]{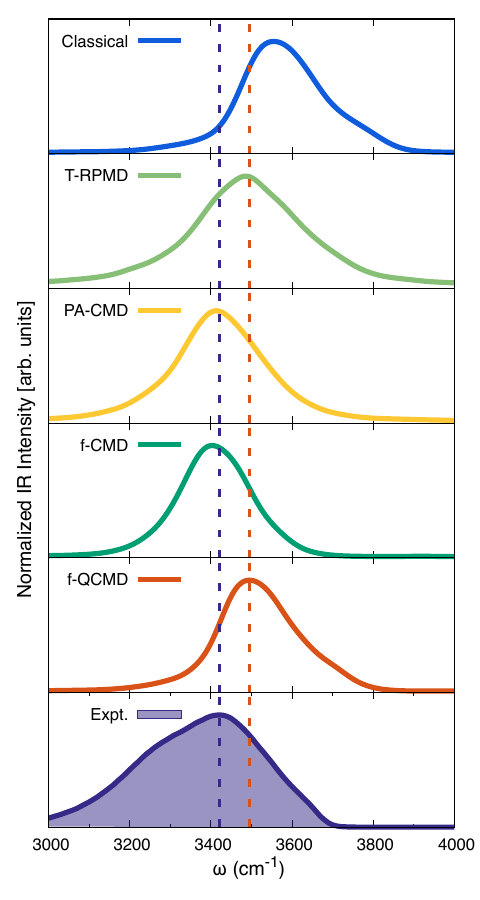}   
    \caption{Calculated IR spectra of the liquid bulk water in the OH stretch region at 300 K using different simulation methods compared to the experiment,\cite{augBertie1996} with purple and red dashed lines representing the experimental and f-QCMD peak positions, respectively. See the SI Figure S5 for the full spectra. Note that the good performance of PA-CMD and f-CMD is an artifact of parameterizing the q-TIP4P/F model to reproduce the experimental spectrum using CMD.~\cite{julHabershon2009} See the text for more details.}
    \label{fig:bulk-water-spec}
\end{figure}
  The simulated spectra for the OH stretch region of the liquid bulk water at 300 K is presented in Figure~\ref{fig:bulk-water-spec}, compared to the experimental spectra from Ref.~\citenum{augBertie1996} (the full spectra can be found in the Supporting Information (SI) Figure S5). While the use of the total dipole derivative as opposed to the molecular dipole moments results in slight differences in the line shapes from those reported previously,\cite{aprLondon2024} the general performances of the methods shown here remain the same. All methods agree with each other reasonably well for the bending and libration bands (SI Figure S5). However, the differences emerge when focusing on the OH stretching band. Classical dynamics exhibits significant blue-shifting of the peak due to the neglect of NQEs. The T-RPMD spectrum, with its inclusion of NQEs, is thus red-shifted and slightly broadened compared to the classical spectra and is in better agreement with the experimental spectra. 
The curvature problem does appear in the PA-CMD spectrum, as evidenced by the red-shifting of the OH peak compared to T-RPMD. However, the PA-CMD spectrum has better agreement with the experiment simply because the q-TIP4P/F model was parameterized to reproduce the experimental spectrum using CMD.~\cite{julHabershon2009} The corresponding spectra are also calculated for the new methods included in DL\_POLY Quantum 2.1. The f-QCMD spectrum demonstrates the method's ability to overcome the curvature problem with its OH peak in line with T-RPMD without artificial broadening. f-CMD can reproduce the PA-CMD spectrum very well, with only slight red-shifting of the OH peak, which has also been observed in bulk water spectra for the force-matching f-CMD method.~\cite{mayYuan2018}

\begin{figure}[!h]
    \centering
    \includegraphics[width=\linewidth]{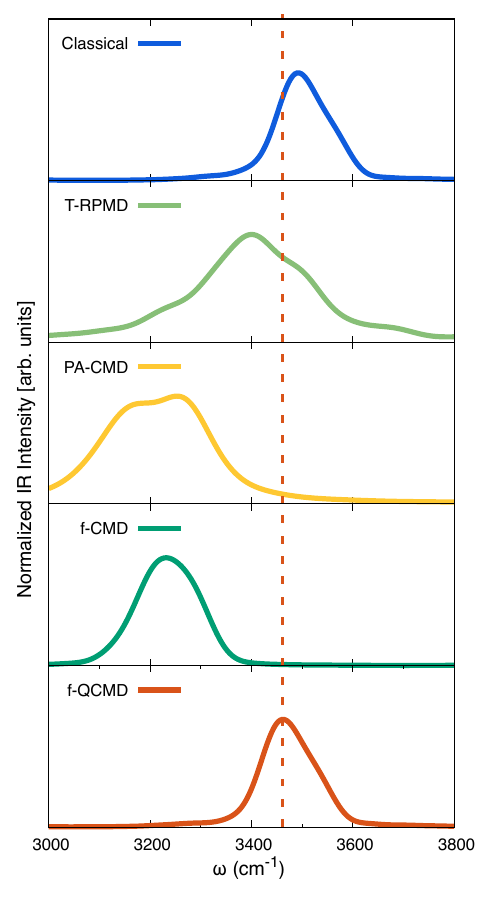}   
    \caption{Calculated IR spectra of ice I$_{\mathrm{h}}$ at 150 K using different methods. Dashed line
    indicates the f-QCMD peak position.}
    \label{fig:bulk-ice-spec}
\end{figure}

To further demonstrate the capabilities of the different PI methods included in the DL\_POLY Quantum 2.1 version, the
spectra for ice I$_{ \mathrm{h}}$ in the OH stretch region is simulated and presented in Figure~\ref{fig:bulk-ice-spec}
with the full spectra given in the SI Figure S10. We do not compare to experiment here as available experimental spectra
are for polycrystalline ice~\cite{augSmit2017} while our simulations are for a single crystal of ice I$_\mathrm{h}$.
Again, we see a blue shift in the simulated classical spectra compared to the PI methods. The more interesting comparison is between the different PI methods, as there is a larger difference amongst their spectra than there is for the liquid bulk water case. We see the exacerbation of the curvature problem as the PA-CMD and f-CMD spectra are further red-shifted from the T-RPMD and f-QCMD spectra when compared to the liquid bulk water at 300 K. The position of the f-QCMD peak is mildly red-shifted compared to the liquid bulk water ($\sim 30$ cm$^{-1}$) and is consistent with the reported spectra for q-TIP4P/F with both adiabatic QCMD~\cite{novTrenins2022} and f-QCMD~\cite{octLawrence2023}. We note that there is a difference in line shape due to our use of a smaller Hann window cut-off, which removes the splitting of the symmetric and antisymmetric stretching peaks that is an artifact of the q-TIP4P/F potential~\cite{augTrenins2019,janWillatt2018}. Additionally, the reported spectra for q-TIP4P/F perform their dynamics simulations in the NVT ensemble with a gentle thermostat instead of the NVE ensemble we use here.~\cite{janWillatt2018,augTrenins2019,decBenson2019} As expected, the T-RPMD spectrum displays significant broadening of the peak with the lowering of the temperature. The peak position is slightly red-shifted compared to f-QCMD, which is not seen in previously reported spectra.~\cite{augTrenins2019} This could be due to our implementation of T-RPMD, which uses the PILE thermostat with a critical damping factor for the internal modes, while those reported are under-damped.~\cite{augTrenins2019} 
\\\\
\textbf{Li-TFSi Aqueous Electrolyte Solutions}
\\\\
\begin{figure*}[!t]
    \centering
    \includegraphics[width=\linewidth]{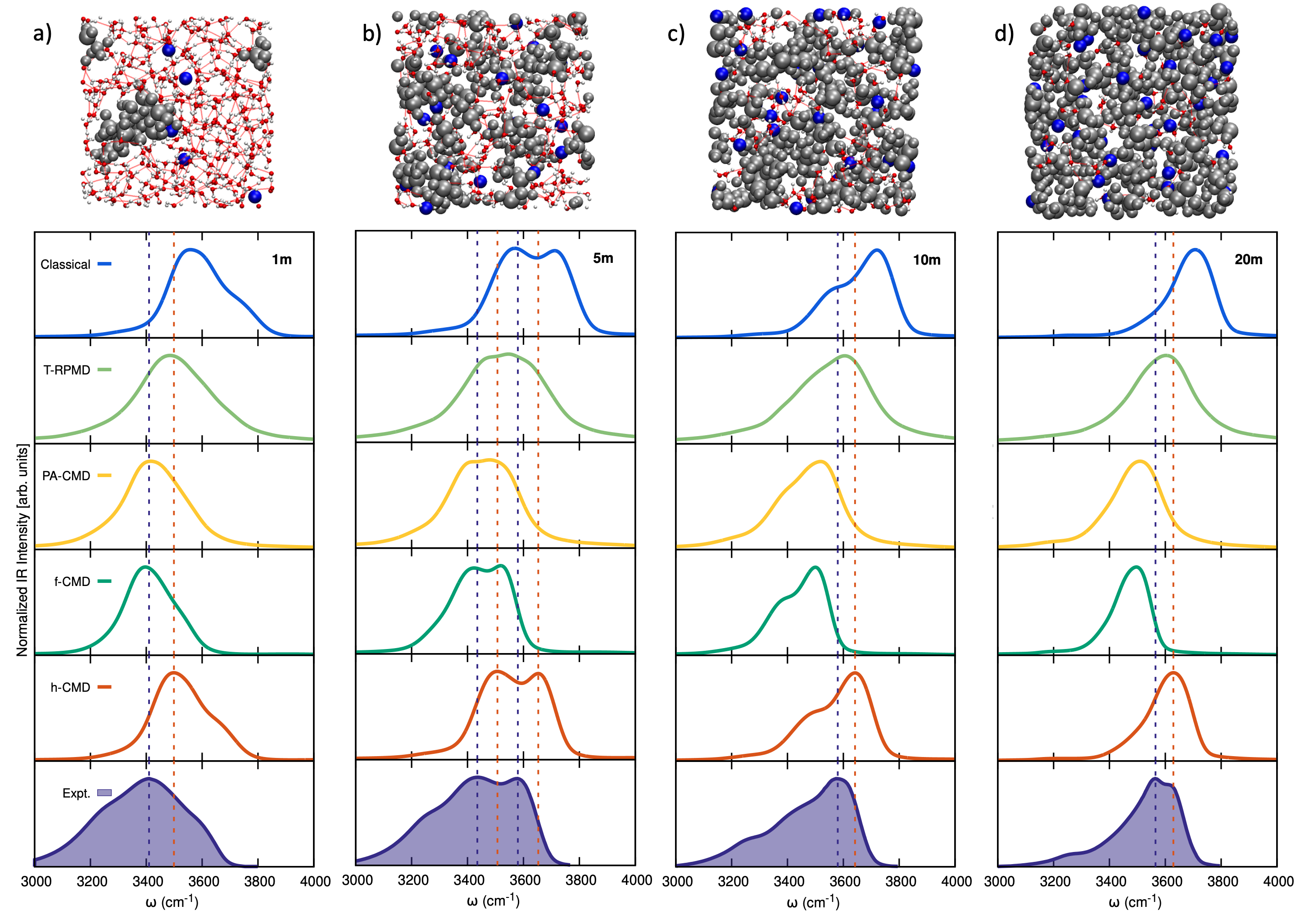}   
    \caption{(a-d) Snapshots of the simulated Li-TFSI aqueous electrolyte solutions with 1-20 m concentrations at 298 K (upper panel). The Li$^+$ ions are in blue, the TFSI$^-$ ions are shown in gray, and the water molecules are in red and white. The lower panel shows calculated IR spectra in the OH stretch region as the function of Li-TFSI concentrations at 298 K using different simulation methods compared to the experiment from Ref.~\citenum{litfsi_spectra}. Purple and red vertical dashed lines represent experimental and h-CMD peak positions, respectively. Note that, similar to the liquid bulk water case discussed above, the good performance of PA-CMD and f-CMD for the dilute 1m concentration, which resembles that of pure liquid bulk water, is an artifact of the q-TIP4P/F water potential model. As expected, this artificial agreement diminishes as the Li-TFSI concentration increases. See the text for more details.}    
    \label{fig:spec-litfsi}
\end{figure*}
The development of water-in-salt (WiS) electrolytes has garnered significant interest due to their expanded electrochemical stability window and potential for stable and reversible electrochemical applications enabling high--voltage aqueous batteries.~\cite{wis_litfsi} Aqueous Li-TFSI, a prominent representative, is one of the candidates that exhibits unique structural and dynamic behaviors at high concentrations. Experimental investigations, particularly using vibrational spectroscopic techniques like IR, have revealed the intricate interplay between solvation structures and ion-ion vs. ion-water interactions in aqueous Li-TFSI solutions.~\cite{litfsi_ir_2018,zhang2020understanding,litfsi_spectra} While classical MD simulations have been extensively used to study the structural properties of WiS electrolytes, they fall short in capturing NQEs, which are essential for accurately representing vibrational spectra. Also, as mentioned for the case of liquid bulk water and ice I$_{ \mathrm{h}}$ systems above and also recently highlighted in our previous work for D$_2$O confined in ZIF-90 framework,\cite{novLimbu2024} the legacy PA-CMD and T-RPMD methods, have well-known shortcomings when applied to simulating the spectral signatures of interfaces. In this case, to evaluate the accuracy of our h-CMD scheme compared to other approximate PI methods, we compare the simulated IR spectra of aqueous Li-TFSI electrolyte solutions with experimental data as a benchmark.

For the Li-TFSI/water system, force field parameters were adopted from Ref.~\citenum{litfsi_spectra}. The TFSI$^-$ was modeled using the generalized Amber force field (GAFF),\cite{GAFF2004} the Li$^+$ parameters were taken from the optimized potentials for liquid simulations (OPLS) force field,\cite{opls1996} and the q-TIP4P/F water model\cite{julHabershon2009} was used to describe water. Li-TFSI--water interactions were included using the non-bonded electrostatic and Lennard-Jones (LJ) potentials. Lorentz--Berthelot mixing rules were used to drive cross-interaction terms. However, this classical force field setup was found to fail to reproduce key experimental features, particularly for the intermediate 5m concentration, where the doublet characteristic was notably absent. Therefore, using the 5m concentration system, the non-bonded LJ parameters between all the Li-TFSI salt atoms and the water oxygen atom were adjusted to align the spectral positions and line shapes with experimental peaks using classical simulations. After extensive benchmarks using the experimental IR spectra as reference (see SI Section S1 for details), the $\epsilon$ and $\sigma$ parameters for TFSI--O$_W$ interactions were increased by 50\% and 5\%, respectively. For the Li--O$_W$ pairwise interactions, the $\sigma$ parameter was decreased by 15\%, while $\epsilon$ remained unchanged. The fine-tuned force field was then applied to all concentrations for calculating the IR spectra using various PI-based methods, including our newly developed h-CMD method.

Figure~\ref{fig:spec-litfsi} presents the IR spectra in the O--H stretch region from salt-in-water (SiW, 1m) to WiS
(20m) for aqueous Li-TFSI electrolyte solutions (for the full spectra, see SI Figures S29-S32). From Figure~\ref{fig:spec-litfsi}(a) for 1m, the classical spectrum shows a pronounced blue shift in the OH stretch peak due to the neglect of NQEs. In contrast, the T-RPMD spectrum incorporates NQEs, resulting in a red-shifted OH peak that aligns more closely with the experiment. PA-CMD spectrum, however, exhibits an even more red-shifted OH peak relative to T-RPMD and shows a better agreement with the experimental spectrum compared to T-RPMD. However, this is artificial due to the way the q-TIP4P/F water model is parameterized. Expectedly, the spectra for the dilute 1m Li-TFSI solution is calculated to be rather similar to that of pure liquid bulk water (see Figures \ref{fig:bulk-water-spec} and \ref{fig:spec-litfsi}). The h-CMD spectrum generates an OH peak that aligns well with T-RPMD, but overcomes the inherent artificial broadening of the T-RPMD method. As expected, the f-CMD spectrum closely replicates the PA-CMD spectrum, with only a slight red shift in the OH peak.

Figure~\ref{fig:spec-litfsi}(b) shows the calculated IR spectra for the 5m Li-TFSI electrolyte solution, with the most striking difference being the presence of a doublet feature in the experimental spectrum. In the classical MD spectrum, the OH stretching peaks are again blue-shifted. While the T-RPMD spectrum reduces this blue shift and positions the peaks closer to the experiment, it fails to resolve the doublet feature accurately, with the spectrum significantly broadened. The PA-CMD spectrum not only struggles with the curvature problem but also fails to capture this distinct doublet feature, further deviating from the experiment. Similarly, the f-CMD spectrum behaves much like PA-CMD. In contrast, the h-CMD method successfully reproduces the doublet feature and aligns the peak positions closer to the T-RPMD, demonstrating its capability to overcome the curvature problem and reproduce the experimental results with higher accuracies. A similar performance of the h-CMD method is observed in reproducing IR spectra of higher concentrations of Li-TFSI electrolyte solutions, as can be seen from Figure~\ref{fig:spec-litfsi}(c)-(d). This shows the superior performance of h-CMD in addressing the limitations of classical and other PI methods while simultaneously preserving the efficiency of calculating vibrational spectra for complex heterogeneous condensed phase and interface systems with thousands of degrees of freedom.
\\\\
\textbf{Approximate h-CMD method for spectral simulations}
\\\\
To further simplify the computational effort needed in calculating IR spectra, we also introduced an approximate version of h-CMD, i.e., h-CMD-H$_2$O, where only inter-molecular correction potentials for water-water interactions were considered in addition to intra-molecular corrections.\cite{novLimbu2024} In other words, by using this approximation, here one can exclude the corrections for water-salt interactions, focusing solely on the three water-water pairs.
\begin{figure}[!h]
\centering
\includegraphics[width=\linewidth]{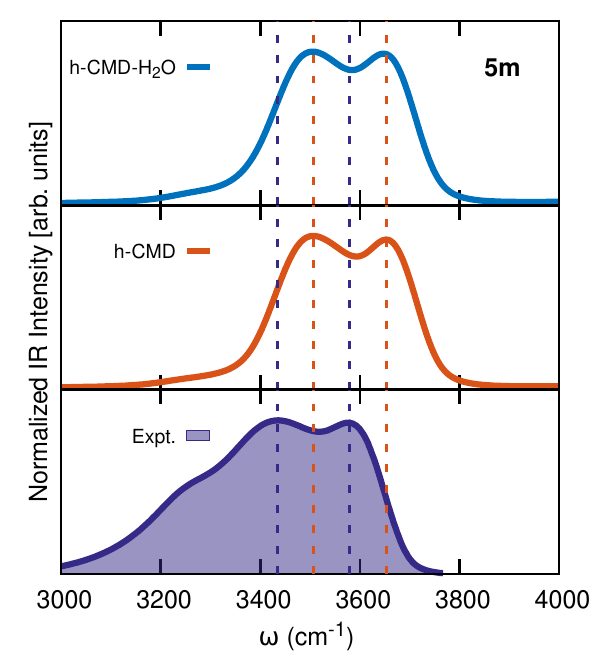}
\caption{Calculated IR spectra of the 5m Li-TFSI electrolyte solution using the approximate h-CMD-H$_2$O method compared to that of full h-CMD and the experiment.\cite{litfsi_spectra}. Purple and red vertical dashed lines represent experimental and h-CMD peak positions, respectively. The spectra for all concentrations are provided in the SI Figure S33.}
\label{spec-hcmd-h2o}
\end{figure}
The resulting IR spectrum for h-CMD-H$_2$O is shown in Figure~\ref{spec-hcmd-h2o} for the 5m Li-TFSI electrolyte solution, alongside the spectrum from the full h-CMD method (which includes 15 pairs of RDF corrections for both water-water and water-salt interactions) and the experiment as a reference. Both h-CMD and h-CMD-H$_2$O produce IR spectra that closely align with that of the experiment.
Also, considering all salt concentrations, the h-CMD-H$_2$O method delivers spectra that are very similar to those of the full h-CMD method (see SI Figure S33), with only a minor blue-shift observed in the higher-frequency peak for the 20m concentration. These findings suggest that the number of correction potentials can be optimized to reduce computational costs significantly without substantial loss of accuracy. This, however, is expected to be system-dependent and needs to be further verified on a case-by-case basis.
\\\\
\section{Conclusions} \label{sec:conclusion}
\textbf{CONCLUSIONS}
\\\\
In this work, we have introduced version 2.1 of our DL\_POLY Quantum software package. Along with some minor bug fixes from previous versions, we have added support for new real-time path integral methods in the form of f-CMD, f-QCMD, and h-CMD. These methods allow for the efficient simulation of vibrational spectra for complex systems with thousands of atoms. DL\_POLY Quantum 2.1 is now the only general-purpose computational software program to include legacy PI methods alongside newly developed ones. We demonstrated the newly included methods, as well as updates to the spectra calculation in this version, through simulations of several systems. For the systems of liquid bulk water and ice I$_\mathrm{h}$, we have shown how different path integral methods are affected by temperature. We also show that our implementation of f-QCMD is able to overcome the curvature problem of PA-CMD and the artificial broadening of T-RPMD. Through calculating the spectra of Li-TFSI systems in aqueous solutions, we have demonstrated the applicability of our recently developed h-CMD method to electrolyte systems and how the DL\_POLY Quantum software can be used to accurately simulate the vibrational spectra of complex, heterogeneous systems with NQEs included.

Our development of DL\_POLY Quantum will continue as we add additional functionalities to the software. The spectra presented in this work highlight the limitations of using analytical force fields. Naturally one would like to move away from them in order to increase the predictive power of the simulations. As such, future versions of the software will include support for dynamics simulations using neural network potentials (NNPs) to evolve the system. The change from analytical force fields to NNPs will also require updates to newly added PI methods, which currently rely on the form of those force fields. These developments are currently underway in our labs.
\\\\
\section{Acknowledgments} \label{sec:acknowledgements}
\textbf{Acknowledgments}
This research was supported by the National Science Foundation through award no. CBET-2302617 and CBET-2302618. Simulations used resources from Bridges2 at Pittsburgh Supercomputing Center through allocation PHY230030P from the Extreme Science and Engineering Discovery Environment (XSEDE),\cite{xsede} which was supported by National Science Foundation grant number 1548562. The use of computing resources and support provided by the HPC center at UMKC is also gratefully acknowledged.
\\\\
\noindent\textbf{Data Availability Statement}
\\All data presented in this work was created using the DL\_POLY Software package which is available from
\\https://github.com/dlpolyquantum/dlpoly\_quantum.
\\The DL\_POLY Quantum documentation is available from 
\\https://dlpolydocs.readthedocs.io/en/latest/

\begin{suppinfo}
Convergence of f-QCMD and f-CMD intra-molecular and radial distribution functions of liquid bulk water at 300K and ice I$_{\mathrm{h}}$ at 150K over IBI iterations compared to the exact PIMD results, details of fine-tuning force field parameters for the intermediate 5m Li-TFSI aqueous electrolyte solution, convergence of the f-CMD intra-molecular bond length and angle distribution functions of the Li-TFSI salts with different concentrations over IBI iterations compared to the exact PIMD results, convergence of the f-CMD radial distribution functions for all Li-TFSI-water pairs with different concentrations over IBI iterations compared to the exact PIMD results, full IR spectra for liquid bulk water, ice I$_{\mathrm{h}}$, and all studied concentrations of Li-TFSI aqueous solutions using different methods, simulated IR spectra using the approximate h-CMD-H$_2$O method compared to the full h-CMD method and the experiment for all Li-TFSI concentrations, and average simulation temperatures of all studied systems with different methods and for all NVE trajectories.
\end{suppinfo}

\bibliography{jpc-soft}

\begin{tocentry}
\includegraphics[scale=0.3]{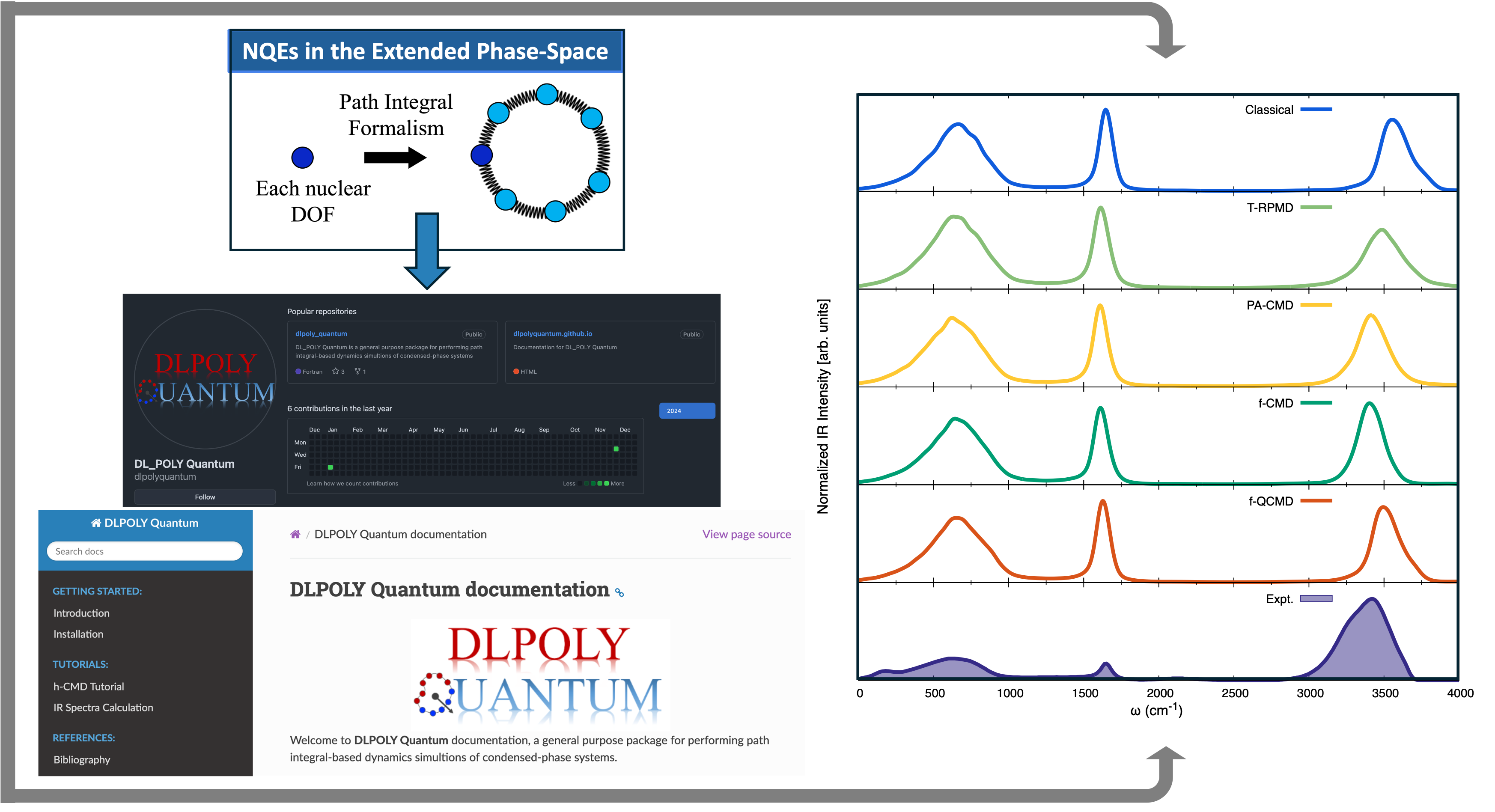}
TOC entry
\end{tocentry}

\end{document}


\newpage
\tableofcontents
\newpage

\clearpage
\begin{figure}[!ht] 
\centering 
\includegraphics[width=\linewidth]{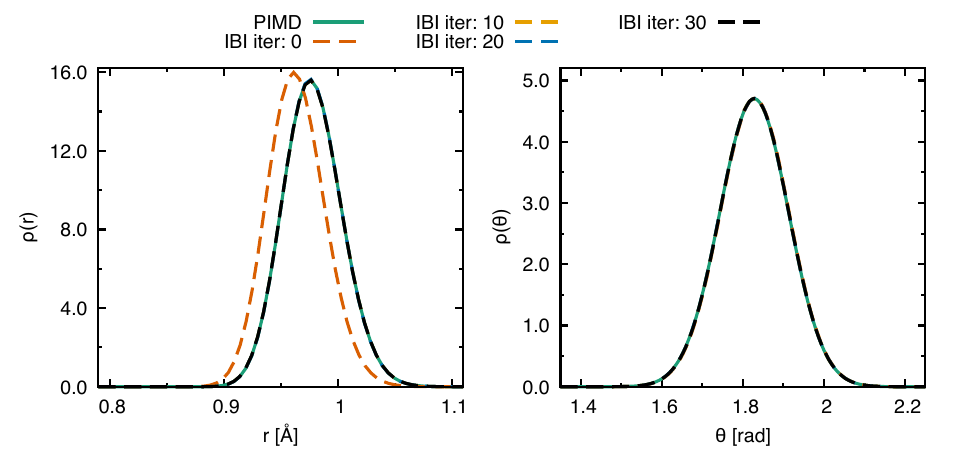} 
    \caption{Convergence of f-QCMD intra-molecular distribution functions for liquid bulk water.}
\label{fig:fq-water-intra}
\addcontentsline{toc}{subsection}{Figure \ref{fig:fq-water-intra}. 
Convergence of f-QCMD intra-molecular distribution functions for liquid bulk water.}
\end{figure}

\clearpage
\begin{figure}[!ht] 
\centering 
\includegraphics[width=\linewidth]{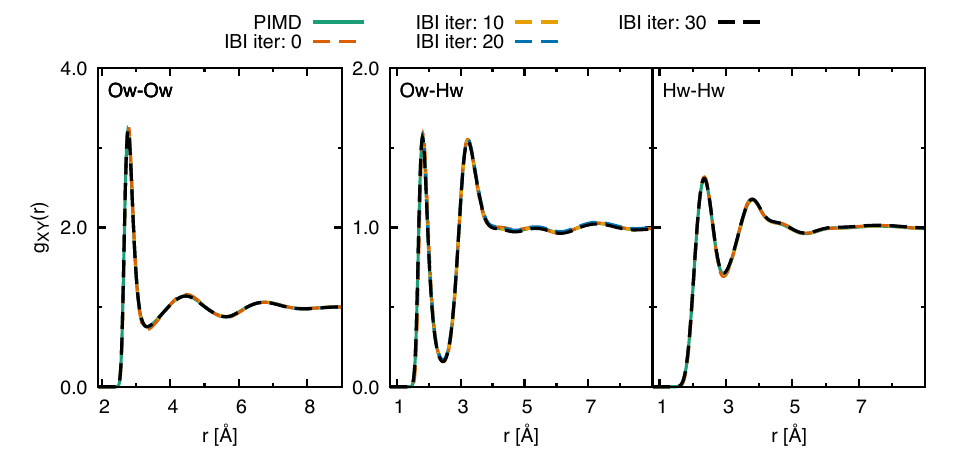} 
    \caption{Convergence of f-QCMD radial distribution functions for liquid bulk water.}
\label{fig:fq-water-rdf}
\addcontentsline{toc}{subsection}{Figure \ref{fig:fq-water-rdf}. 
Convergence of f-QCMD radial distribution functions for liquid bulk water.}
\end{figure}

\clearpage
\begin{figure}[!ht] 
\centering 
\includegraphics[width=\linewidth]{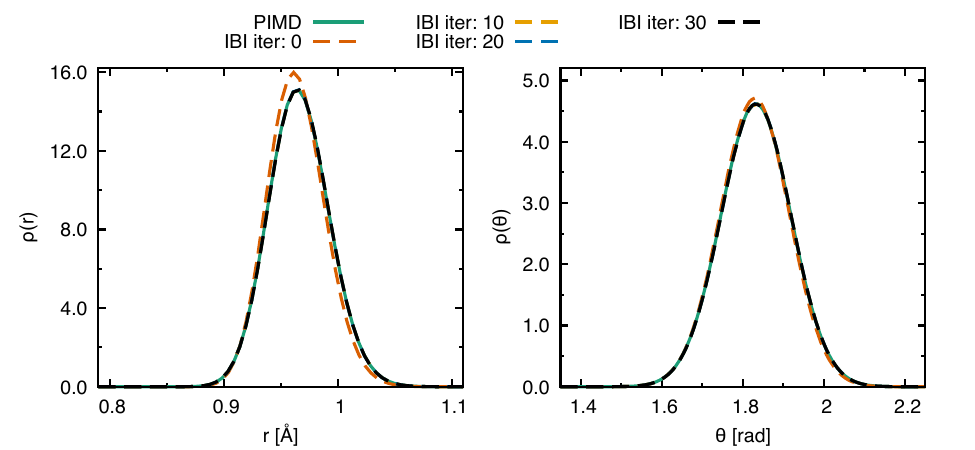} 
    \caption{Convergence of f-CMD intra-molecular distribution functions for liquid bulk water.}
\label{fig:fc-water-intra}
\addcontentsline{toc}{subsection}{Figure \ref{fig:fc-water-intra}. 
Convergence of f-CMD intra-molecular distribution functions for liquid bulk water.}
\end{figure}

\clearpage
\begin{figure}[!ht] 
\centering 
\includegraphics[width=\linewidth]{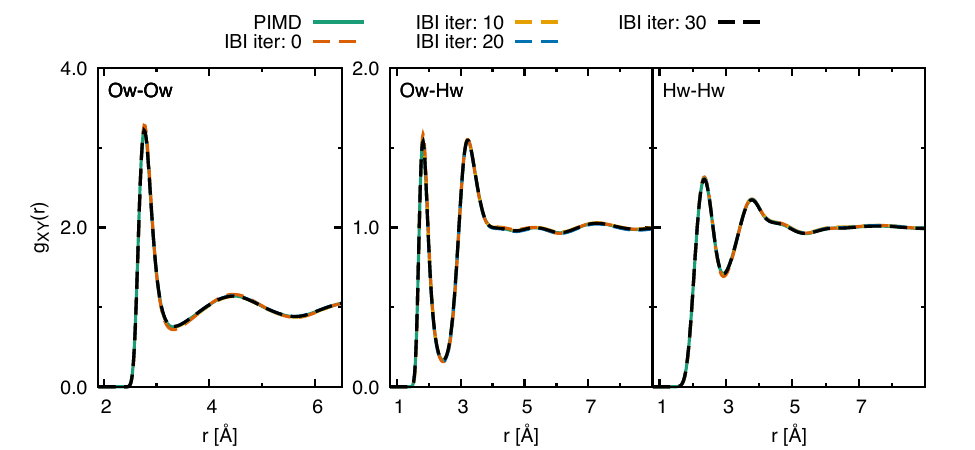} 
    \caption{Convergence of f-CMD radial distribution functions for liquid bulk water.}
\label{fig:fc-water-rdf}
\addcontentsline{toc}{subsection}{Figure \ref{fig:fc-water-rdf}. 
Convergence of f-CMD RDFs for liquid bulk water.}
\end{figure}

\clearpage
\begin{figure}[!ht] 
\centering 
\includegraphics[width=\linewidth]{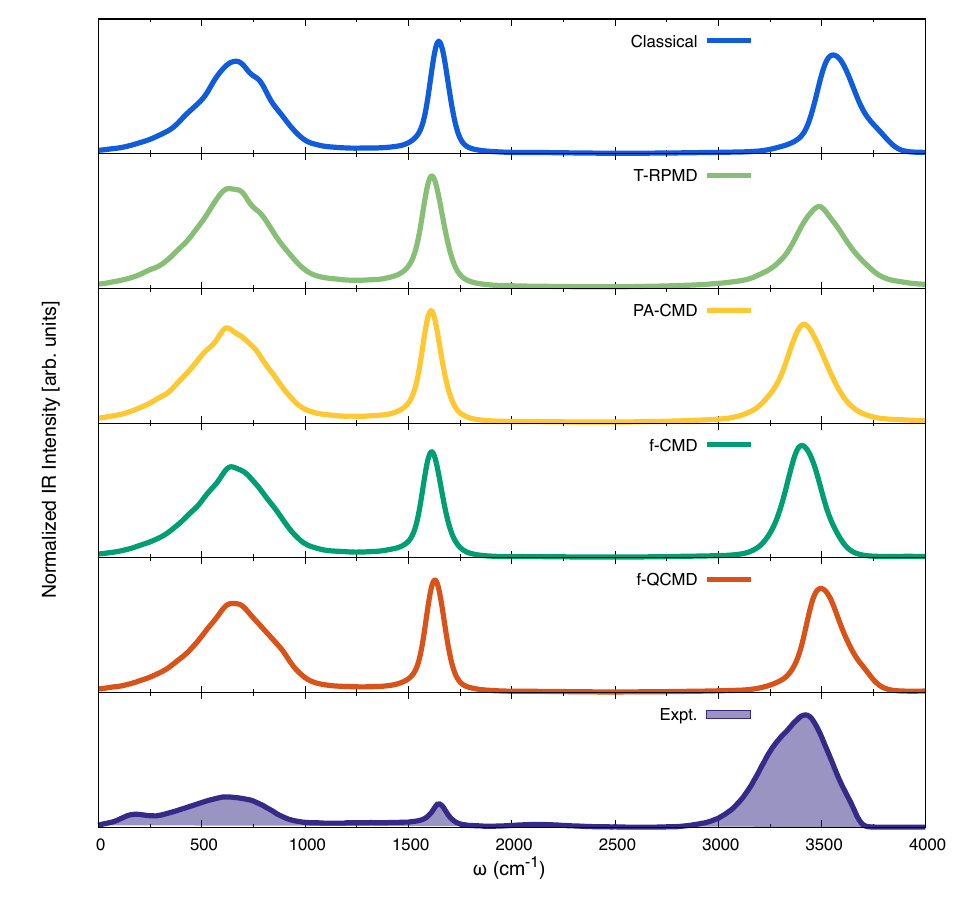} 
    \caption{Calculated full IR spectra for liquid bulk water at 300 K using different simulation methods compared to the
    experimental spectrum from Ref.~\citenum{augBertie1996}.}
\label{fig:spec-water}
\addcontentsline{toc}{subsection}{Figure \ref{fig:spec-water}. Full IR spectra for liquid bulk water at 300 K.}
\end{figure}

\clearpage
\begin{table}
\caption{Average simulation temperature of the liquid bulk water system with different simulation methods for all 50 NVE trajectories.}
\addcontentsline{toc}{subsection}{Table \ref{tab:water-avg-t}. Average simulation temperature of the liquid bulk water system.}
\centering
\setlength{\tabcolsep}{4mm}
\scalebox{0.8}{
\begin{tabular}{c c c c c c}
\midrule[1.4pt]
Trajectory & MD  & T-RPMD  & PA-CMD & f-QCMD & f-CMD  \\
\midrule[1.4pt]
  1 &  296.8 &  299.8 &  300.5 &  301.9 &  299.2 \\
  2 &  298.7 &  299.8 &  300.6 &  302.0 &  300.4 \\
  3 &  295.4 &  299.9 &  300.6 &  302.0 &  303.7 \\
  4 &  297.9 &  299.8 &  300.6 &  302.6 &  299.1 \\
  5 &  300.2 &  300.0 &  300.5 &  308.7 &  298.1 \\
  6 &  299.1 &  299.0 &  300.4 &  301.1 &  302.9 \\
  7 &  297.8 &  299.9 &  300.7 &  302.0 &  300.6 \\
  8 &  296.8 &  299.9 &  300.5 &  298.8 &  302.6 \\
  9 &  293.0 &  299.9 &  300.6 &  296.4 &  301.2 \\
 10 &  295.4 &  299.8 &  300.6 &  303.5 &  304.5 \\
 11 &  299.4 &  299.8 &  300.5 &  303.9 &  307.1 \\
 12 &  300.7 &  299.9 &  300.5 &  296.4 &  297.6 \\
 13 &  304.4 &  299.9 &  300.6 &  297.6 &  298.7 \\
 14 &  304.2 &  299.9 &  300.5 &  302.3 &  298.3 \\
 15 &  303.9 &  299.8 &  300.6 &  297.3 &  297.1 \\
 16 &  309.9 &  299.8 &  300.6 &  301.6 &  300.9 \\
 17 &  299.5 &  299.8 &  300.6 &  301.1 &  297.6 \\
 18 &  294.1 &  299.9 &  300.4 &  297.7 &  301.4 \\
 19 &  299.1 &  300.0 &  300.6 &  295.3 &  298.0 \\
 20 &  294.3 &  299.9 &  300.6 &  296.8 &  294.3 \\
 21 &  293.7 &  299.8 &  300.6 &  298.5 &  296.5 \\
 22 &  297.8 &  299.8 &  300.7 &  292.5 &  291.2 \\
 23 &  289.9 &  299.9 &  300.5 &  298.0 &  303.0 \\
 24 &  292.9 &  300.0 &  300.7 &  300.2 &  297.4 \\
 25 &  292.5 &  299.9 &  300.7 &  303.8 &  296.4 \\
 26 &  299.4 &  299.9 &  300.7 &  298.8 &  304.5 \\
 27 &  298.0 &  299.9 &  300.5 &  295.5 &  301.8 \\
 28 &  297.6 &  300.0 &  300.6 &  299.8 &  302.9 \\
 29 &  298.8 &  299.0 &  300.6 &  298.3 &  306.6 \\
 30 &  298.7 &  299.9 &  300.5 &  296.9 &  295.6 \\
 31 &  298.6 &  299.9 &  300.6 &  298.9 &  293.4 \\
 32 &  299.4 &  299.8 &  300.6 &  298.5 &  300.6 \\
 33 &  298.1 &  299.9 &  300.5 &  298.9 &  301.8 \\
 34 &  298.8 &  299.9 &  300.7 &  297.8 &  299.8 \\
 35 &  300.0 &  299.9 &  300.5 &  297.0 &  301.2 \\
 36 &  294.6 &  299.9 &  300.4 &  300.7 &  302.7 \\
 37 &  294.4 &  299.9 &  300.6 &  293.8 &  300.6 \\
 38 &  301.2 &  299.8 &  300.5 &  295.4 &  300.4 \\
 39 &  304.5 &  299.9 &  300.5 &  298.6 &  303.4 \\
 40 &  300.4 &  299.9 &  300.6 &  297.9 &  302.0 \\
 41 &  298.3 &  299.8 &  300.4 &  299.4 &  300.1 \\
 42 &  300.6 &  299.9 &  300.4 &  294.6 &  299.4 \\
 43 &  299.0 &  299.8 &  300.6 &  296.1 &  297.4 \\
 44 &  298.7 &  299.8 &  300.5 &  294.8 &  297.2 \\
 45 &  301.0 &  299.9 &  300.6 &  300.7 &  297.5 \\
 46 &  295.1 &  299.9 &  300.5 &  299.9 &  302.9 \\
 47 &  298.0 &  299.9 &  300.7 &  301.6 &  306.2 \\
 48 &  298.2 &  299.8 &  300.5 &  304.2 &  298.6 \\
 49 &  303.2 &  299.8 &  300.5 &  298.4 &  299.1 \\
 50 &  296.8 &  299.8 &  300.6 &  304.4 &  302.0 \\
 \hline
Average & 298.4 & 299.8 & 300.6 & 299.3 & 300.1 \\
\midrule[1.4pt]
\end{tabular}}     
\label{tab:water-avg-t}
\end{table}

\clearpage
\begin{figure}[!ht] 
\centering 
\includegraphics[width=\linewidth]{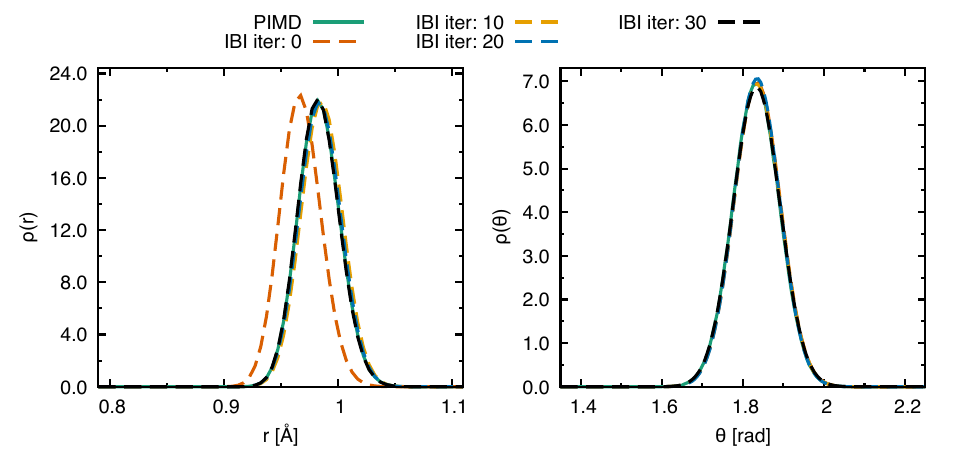} 
    \caption{Convergence of f-QCMD intra-molecular distribution functions for ice I$_\mathrm{h}$.}
\label{fig:fq-ice-intra}
\addcontentsline{toc}{subsection}{Figure \ref{fig:fq-ice-intra}. 
Convergence of f-QCMD intra-molecular distribution functions for ice I$_\mathrm{h}$.}
\end{figure}

\clearpage
\begin{figure}[!ht] 
\centering 
\includegraphics[width=\linewidth]{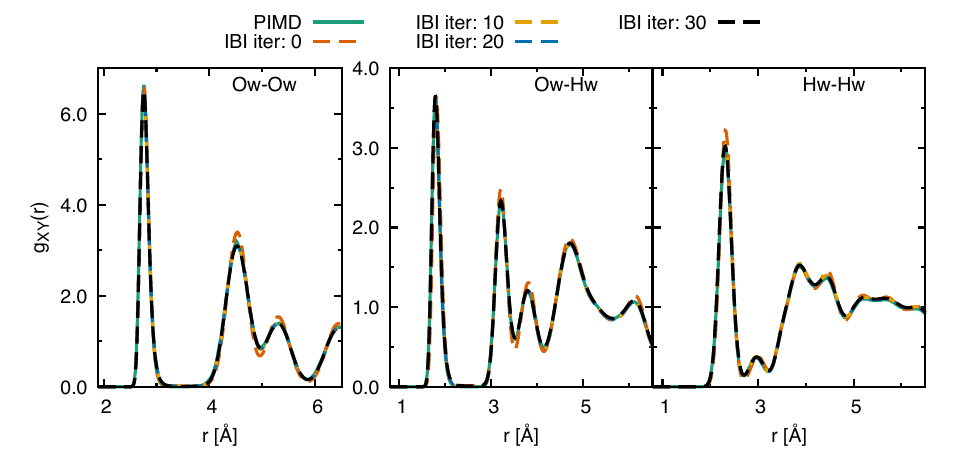} 
    \caption{Convergence of f-QCMD radial distribution functions for ice I$_\mathrm{h}$.}
\label{fig:fq-ice-rdf}
\addcontentsline{toc}{subsection}{Figure \ref{fig:fq-ice-rdf}. 
Convergence of f-QCMD RDFs for ice I$_\mathrm{h}$.}
\end{figure}

\clearpage
\begin{figure}[!ht] 
\centering 
\includegraphics[width=\linewidth]{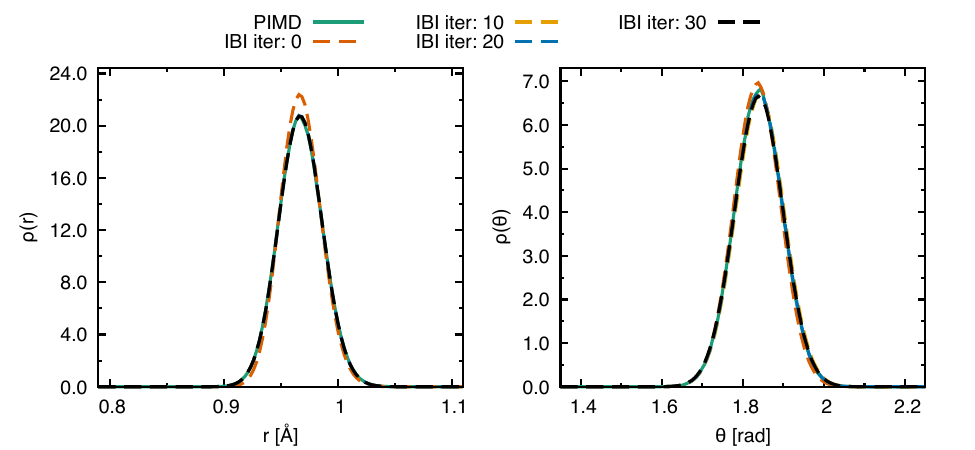} 
    \caption{Convergence of f-CMD intra-molecular distribution functions for ice I$_\mathrm{h}$.}
\label{fig:fc-ice-intra}
\addcontentsline{toc}{subsection}{Figure \ref{fig:fc-ice-intra}. 
Convergence of f-CMD intra-molecular distribution functions for ice I$_\mathrm{h}$.}
\end{figure}

\clearpage
\begin{figure}[!ht] 
\centering 
\includegraphics[width=\linewidth]{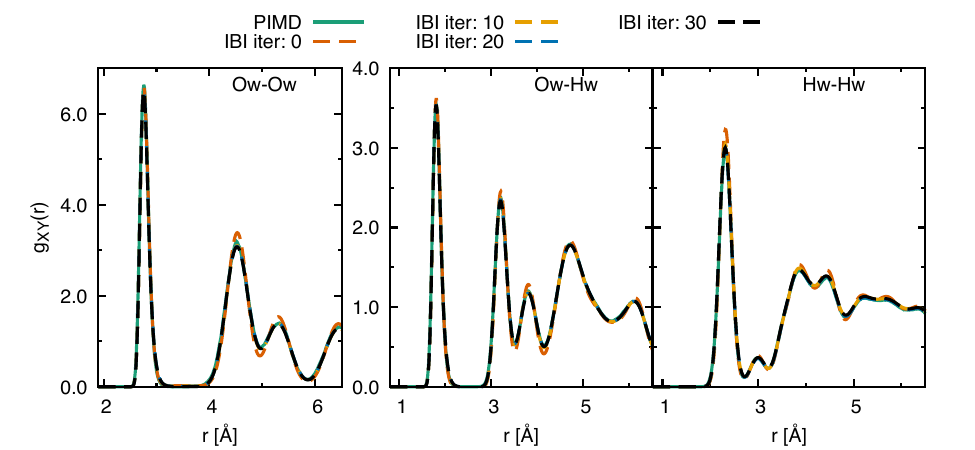} 
    \caption{Convergence of f-CMD radial distribution functions for ice I$_\mathrm{h}$.}
\label{fig:fc-ice-rdf}
\addcontentsline{toc}{subsection}{Figure \ref{fig:fc-ice-rdf}. 
Convergence of f-CMD RDFs for ice I$_\mathrm{h}$.}
\end{figure}

\clearpage
\begin{figure}[!ht] 
\centering 
\includegraphics[width=\linewidth]{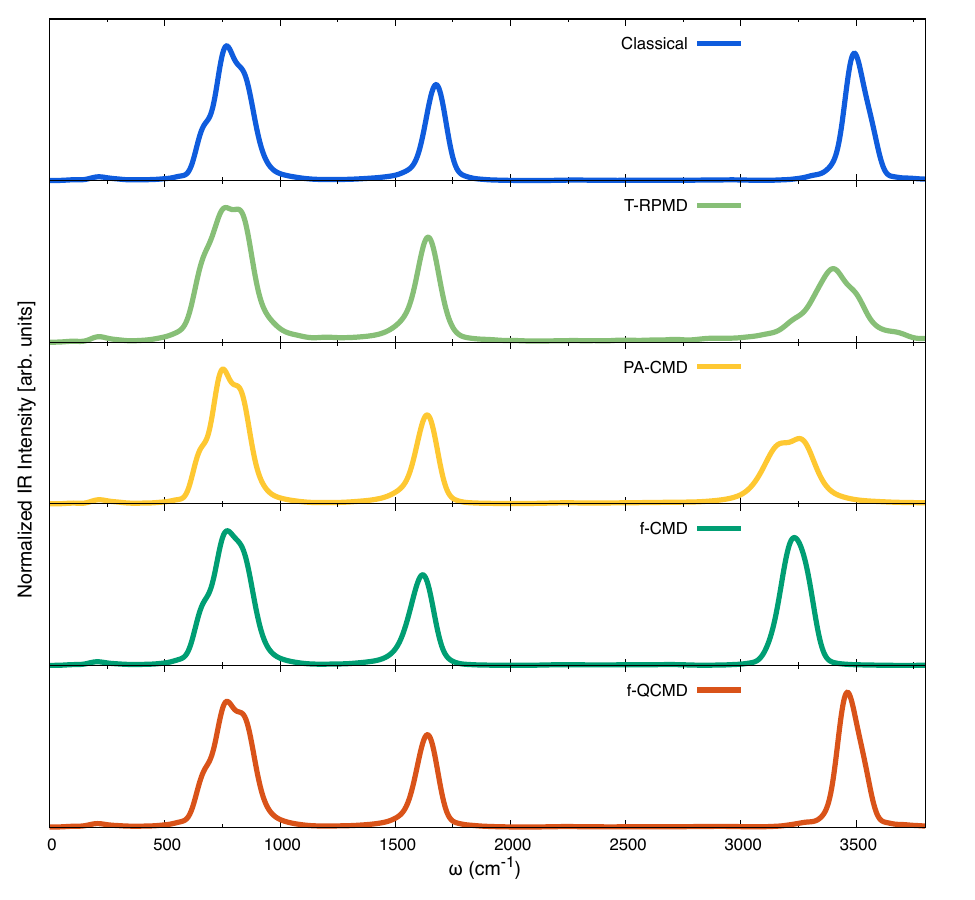} 
    \caption{Calculated full IR spectra for ice I$_\mathrm{h}$ at 150 K using different simulation methods.}
\label{fig:spec-ice}
\addcontentsline{toc}{subsection}{Figure \ref{fig:spec-ice}. Full IR spectra for ice I$_\mathrm{h}$.}
\end{figure}

\clearpage
\begin{table}
  \caption{Average simulation temperature of ice I$_\mathrm{h}$ system with different simulation methods for all 50 NVE trajectories.}
\addcontentsline{toc}{subsection}{Table \ref{tab:ice-avg-t}. Average simulation temperature of ice I$_\mathrm{h}$ system.}
\centering
\setlength{\tabcolsep}{4mm}
\scalebox{0.8}{
\begin{tabular}{c c c c c c}
\midrule[1.4pt]
Trajectory & MD  & T-RPMD  & PA-CMD & f-QCMD & f-CMD  \\
\midrule[1.4pt]
  1 &  150.0 &  150.0 &  150.0 &  149.8 &  149.8 \\
  2 &  149.9 &  150.0 &  150.0 &  151.9 &  149.6 \\
  3 &  150.4 &  150.0 &  149.9 &  150.6 &  149.1 \\
  4 &  149.8 &  149.9 &  149.9 &  151.7 &  151.1 \\
  5 &  150.8 &  150.0 &  149.9 &  148.8 &  148.5 \\
  6 &  149.3 &  150.0 &  149.9 &  150.7 &  148.4 \\
  7 &  150.8 &  150.0 &  150.0 &  149.8 &  150.6 \\
  8 &  149.4 &  150.0 &  149.9 &  150.3 &  149.6 \\
  9 &  150.5 &  150.0 &  149.9 &  150.8 &  150.9 \\
 10 &  149.0 &  150.0 &  150.0 &  150.4 &  151.5 \\
 11 &  150.8 &  150.0 &  149.9 &  149.5 &  151.2 \\
 12 &  149.5 &  150.0 &  150.0 &  150.1 &  149.9 \\
 13 &  150.3 &  150.0 &  150.0 &  150.6 &  149.5 \\
 14 &  149.7 &  150.0 &  150.0 &  149.8 &  150.8 \\
 15 &  150.4 &  150.0 &  150.0 &  150.0 &  151.7 \\
 16 &  149.7 &  150.0 &  150.0 &  151.4 &  151.4 \\
 17 &  149.7 &  150.0 &  149.9 &  148.3 &  149.5 \\
 18 &  151.0 &  150.0 &  150.0 &  150.0 &  151.6 \\
 19 &  149.8 &  150.0 &  150.0 &  151.7 &  149.7 \\
 20 &  150.2 &  150.0 &  149.9 &  150.8 &  148.5 \\
 21 &  149.7 &  150.0 &  150.0 &  149.1 &  150.0 \\
 22 &  149.8 &  150.0 &  150.0 &  151.2 &  151.2 \\
 23 &  150.2 &  150.0 &  150.0 &  150.3 &  150.1 \\
 24 &  149.4 &  150.0 &  150.0 &  150.8 &  150.4 \\
 25 &  150.4 &  150.0 &  149.9 &  150.3 &  150.6 \\
 26 &  149.5 &  150.0 &  149.9 &  148.4 &  149.2 \\
 27 &  150.7 &  150.0 &  150.0 &  149.3 &  150.8 \\
 28 &  149.9 &  150.0 &  150.0 &  150.7 &  149.1 \\
 29 &  149.9 &  150.0 &  150.0 &  149.5 &  150.1 \\
 30 &  150.6 &  150.0 &  150.0 &  149.1 &  150.9 \\
 31 &  149.7 &  150.0 &  150.0 &  148.0 &  147.4 \\
 32 &  150.5 &  150.0 &  150.0 &  149.2 &  150.8 \\
 33 &  149.4 &  150.0 &  150.0 &  148.6 &  149.9 \\
 34 &  150.3 &  150.0 &  150.0 &  149.7 &  149.4 \\
 35 &  149.6 &  150.0 &  149.9 &  149.9 &  150.0 \\
 36 &  150.7 &  150.0 &  150.0 &  150.9 &  149.8 \\
 37 &  149.3 &  150.0 &  149.9 &  150.5 &  149.8 \\
 38 &  150.9 &  150.0 &  149.9 &  150.7 &  151.0 \\
 39 &  149.3 &  150.0 &  149.9 &  149.9 &  150.5 \\
 40 &  150.9 &  150.0 &  149.9 &  151.0 &  150.4 \\
 41 &  149.7 &  150.0 &  149.9 &  150.5 &  148.3 \\
 42 &  150.1 &  150.0 &  150.0 &  151.0 &  149.6 \\
 43 &  149.4 &  150.0 &  150.0 &  150.8 &  149.4 \\
 44 &  149.9 &  150.0 &  150.0 &  149.3 &  151.5 \\
 45 &  150.0 &  150.0 &  150.0 &  150.9 &  149.8 \\
 46 &  150.4 &  150.0 &  150.0 &  149.8 &  149.2 \\
 47 &  149.6 &  150.0 &  149.9 &  148.5 &  150.2 \\
 48 &  150.3 &  150.0 &  150.0 &  149.2 &  150.8 \\
 49 &  149.7 &  150.0 &  150.0 &  148.6 &  149.1 \\
 50 &  150.0 &  150.0 &  150.0 &  149.6 &  149.9 \\
 \hline
Average & 150.0 & 150.0 & 150.0 & 150.0 & 150.0 \\
\midrule[1.4pt]
\end{tabular}}     
\label{tab:ice-avg-t}
\end{table}

\clearpage
\noindent \textbf{Section S1. Fine-tuning force field parameters for Li-TFSI aqueous solutions}
\addcontentsline{toc}{section}{Section S1. Fine-tuning force field parameters for Li-TFSI aqueous solutions}

\begin{figure}[!h] 
\centering 
\includegraphics[width=0.48\textwidth]{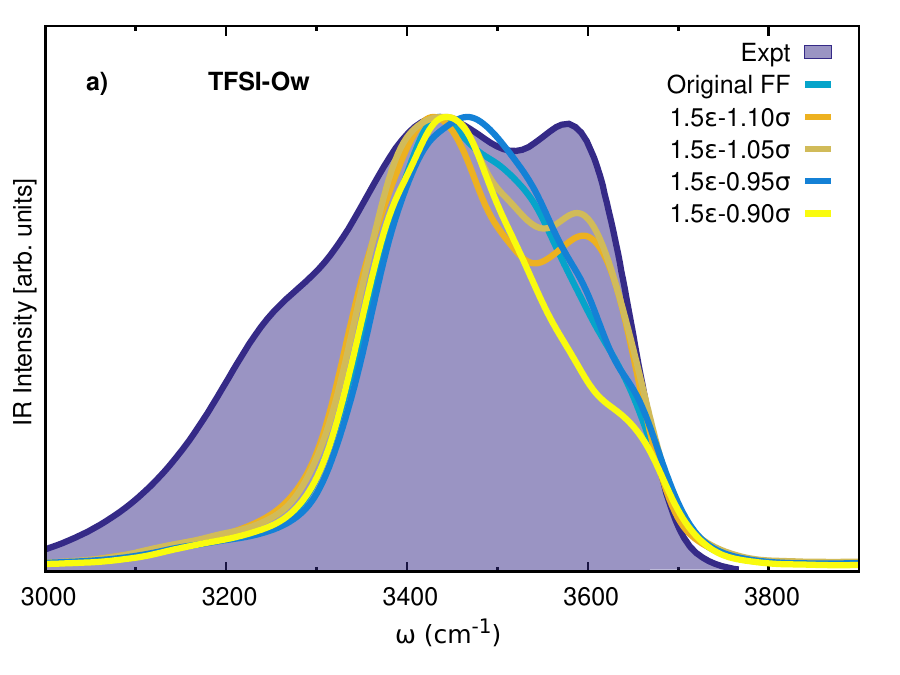} 
\includegraphics[width=0.48\textwidth]{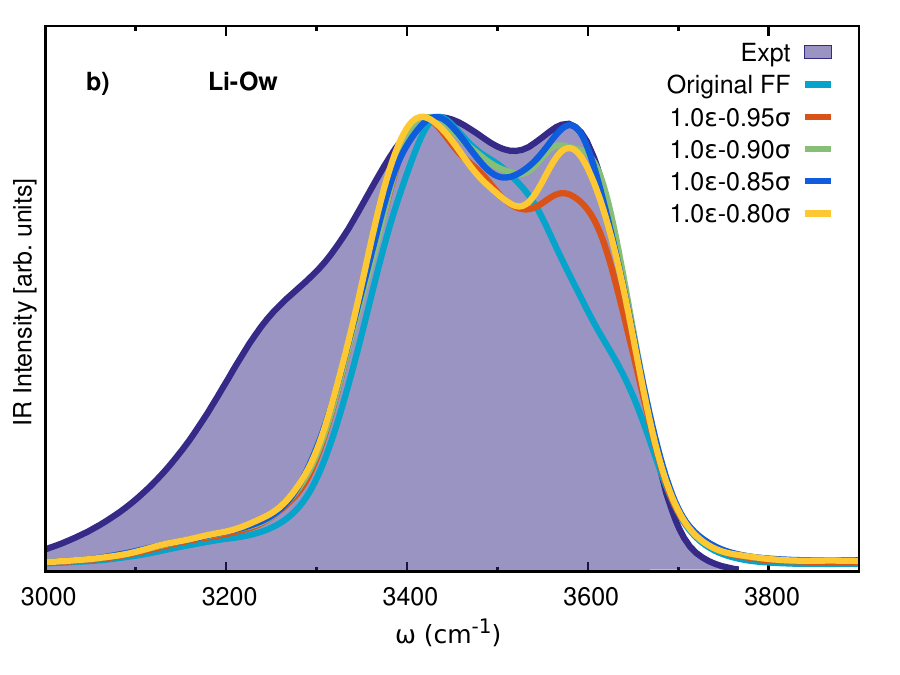}
\caption{Influence of modified force field parameters on the classical MD simulated IR line shapes in the OH stretch region for the \textbf{5m} aqueous Li-TFSI solution. (a) Variations in lineshape of IR spectra resulting from different values of $\sigma$ for 50\% increase in $\epsilon$ for all TFSI--O$_W$ pair interactions. (b) Variations in lineshape of IR spectra for different $\sigma$ values for Li--O$_W$ pair interaction. TFSI-O$_W$ interaction for $\epsilon$ and $\sigma$ are set to increase by 50\%  and 5\%, respectively. All spectra are red-shifted by 137 cm$^{-1}$.}
\label{fig-fit}
\addcontentsline{toc}{subsection}{Figure \ref{fig-fit}. Fine-tuning force field.}
\end{figure}

Accurately simulating the IR spectra of water in heterogeneous environments is challenging, particularly in capturing the complex water-water vs. water-solute interactions. Our initial simulations performed using the original FF showed significant deviations from the experimental spectra, particularly in the O-H stretching region, where the critical doublet feature was absent. This discrepancy underscored the inadequate representation of Li-TFSI--water interactions in the original FF. In this study, non-bonded Lennard-Jones (LJ) potential parameters were modified to better represent these interactions. To systematically understand the effects of parameter variations, we explored how changes in $\epsilon$ and $\sigma$ cross-term parameters influenced the IR line shapes for the intermediate \textbf{5m} Li-TFSI solution based on classical MD simulated IR spectra. As shown in Figure~\ref{fig-fit}(a), modifying $\epsilon$ and $\sigma$ for the TFSI--O$_W$ pairs only initiated the appearance of doublet shoulders; however, the relative intensity in the higher-frequency region remained small, indicating inaccuracies in modeling water-salt interactions. To address this, further refinements were made, with $\epsilon$ for TFSI--O$_W$ interactions increased by 50\% and $\sigma$ by 5\%. The $\sigma$ parameter for Li--O$_W$ interactions was decreased by 15\% with $\epsilon$ remaining unchanged. The resulting spectrum (Figure~\ref{fig-fit}(b)) revealed two distinct peaks with correct relative intensities that matched experimental observations. These features, which were previously underrepresented, were accurately captured after the FF modifications, demonstrating the importance of these adjustments in improving the fidelity of simulated IR spectra for heterogeneous aqueous systems. This fine-tuned FF using classical simulations was then applied across all Li-TFSI concentrations for IR spectral calculations using various path-integral methods, including the newly developed h-CMD method.\cite{novLimbu2024}

\clearpage

\clearpage
\begin{table}
\caption{Force field parameters for Li-TFSI.}
\addcontentsline{toc}{subsection}{Table \ref{table_ff}. Force field parameters for  Li-TFSI.}
    \begin{subtable}{0.49\textwidth}
    \centering
    \caption{\textbf{Non-bonded} terms.}
    \scalebox{0.85}{
    \begin{tabular}{c c c c c}
       \hline
       \specialcell{Atom\\name} & \specialcell{Atom\\type} & \specialcell{q\\(e)} & \specialcell{$\epsilon$\\(kcal/mol)} & \specialcell{$\sigma/2$\\({\AA})} \\
       \hline
        C     & C   &  0.319  & 0.110  & 1.700   \\
        N     & N   & -0.588  & 0.170  & 1.625   \\
        O     & O   & -0.461  & 0.210  & 1.480   \\
        F     & F   & -0.138  & 0.061  & 1.559   \\
        S     & S   & 0.911   & 0.250  & 1.782   \\
        Li    & Li  & 0.800   & 0.018  & 1.06322   \\
        \hline
        \end{tabular}}
    \end{subtable}%
    \begin{subtable}{0.49\textwidth}
    \centering
        \caption{\textbf{Bond length} terms: $U(r) = \frac{1}{2}K_r(r-r_0)^2$}
        \scalebox{0.85}{
        \begin{tabular}{c c c}
        \hline
        \textbf{Bonds} & \specialcell{$K_r$ \\ (kcal.mol$^{-1}${\AA}$^2$)} & \specialcell{r$_0$ \\ ({\AA})} \\
        \hline
        O--S & 985.9928  &  1.466 \\
        C--F & 727.5944  &  1.344 \\
        C--S & 497.7954  &  1.782 \\
        N--S & 514.1978  &  1.752  \\
        \hline
        \end{tabular}}
    \end{subtable}
    \medskip

    \begin{subtable}{0.49\textwidth}
    \centering
    \caption{\textbf{Bond angle} terms: $U(\theta) = \frac{1}{2}K_\theta(\theta-\theta_0)^2$}
    \scalebox{0.85}{
    \begin{tabular}{c c c}
        \hline
        \textbf{Angles} & \specialcell{$K_\theta$ \\ (kcal.mol$^{-1}$.rad$^2$)} & \specialcell{$\theta_0$ \\ ( \degree~)}\\
        \hline
         C--S--O  & 82.5598   &  108.48 \\
         O--S--O  & 90.6000   &  121.88 \\
         N--S--O  & 86.2004   &  107.06 \\
         F--C--S  & 162.4270  &  109.67 \\
         F--C--F  & 142.5174  &  107.16 \\
         C--S--N  &  78.8398  &  103.12 \\
         S--N--S  & 219.4380  &  119.18 \\
        \hline
    \end{tabular}} 
    \end{subtable}%
    \begin{subtable}{0.49\textwidth}
    \centering
        \caption{\textbf{Torsion} terms: $U(\phi) = K_\phi[1+\cos(n\phi-\phi_0)]$}
        \scalebox{0.85}{
        \begin{tabular}{c c c c}
        \hline
        \textbf{Dihedrals} & \specialcell{$K_\phi$ \\ (kcal.mol$^{-1}$)} & \specialcell{$\phi_0$ \\ ( \degree~)} & \specialcell{n \\ ~}\\
        \hline
        F--C--S--O     & 0.143992   &    0  &  3 \\
        S--N--S--O     & 0.499997   &  180  &  3 \\
        S--N--S--O     & 6.799951   &  180  &  1 \\
        F--C--S--N     & 0.143992   &    0  &  3 \\
        S--N--S--C     & 0.499997   &  180  &  3 \\
        S--N--S--C     & 6.799951   &  180  &  1 \\        
        \hline
        \end{tabular}}
    \end{subtable}
    \medskip

    \begin{subtable}{0.45\textwidth}
    \centering
    \caption{\textbf{Fine-tuned non-bonded} LJ cross-term parameters for Li-TFSI and water.}
    \scalebox{0.85}{
    \begin{tabular}{c c c}
       \hline
        \specialcell{Atom\\type pair} & \specialcell{$\epsilon$\\(kcal/mol)} & \specialcell{$\sigma$\\({\AA})} \\
       \hline
        C--O$_W$   &   0.214096  &  3.443423 \\
        N--O$_W$   &   0.266155  &  3.364673 \\
        O--O$_W$   &   0.295816  &  3.212423 \\
        F--O$_W$   &   0.159432  &  3.295373 \\
        S--O$_W$   &   0.322761  &  3.529523 \\
        Li--O$_W$  &   0.057737  &  2.246269 \\
        \hline
        \end{tabular}}
    \end{subtable}%
    \label{table_ff}
\end{table}

\clearpage
\begin{figure}[!ht] 
\centering 
\includegraphics[width=0.99\textwidth]{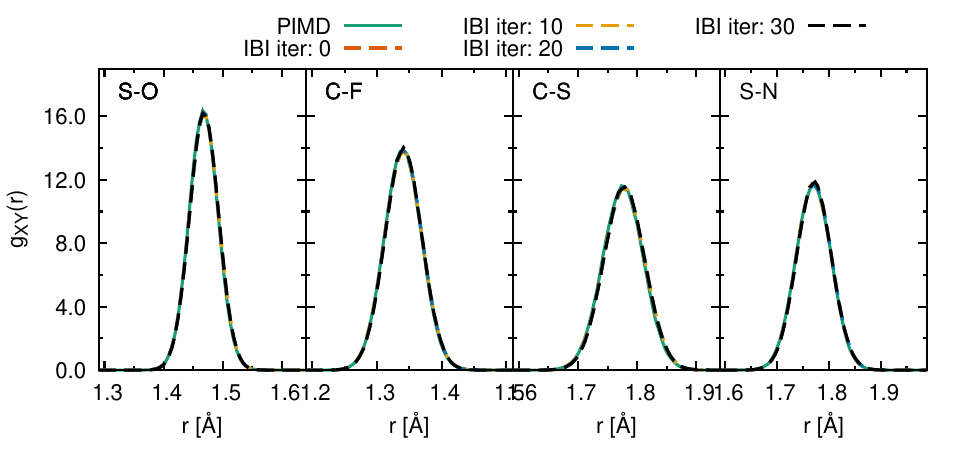} 
\caption{Convergence of f-CMD intra-molecular bond lengths of LiTSFI salt of \textbf{1m} concentration across IBI iterations compared to the exact PIMD results.}
\label{fig-litfsi-bond-1m}
\addcontentsline{toc}{subsection}{Figure \ref{fig-litfsi-bond-1m}. Convergence of f-CMD intra-molecular bond lengths of Li-TFSI salt of \textbf{1m} concentration across IBI iterations. }
\end{figure}

\clearpage
\begin{figure}[!ht] 
\centering 
\includegraphics[width=0.99\textwidth]{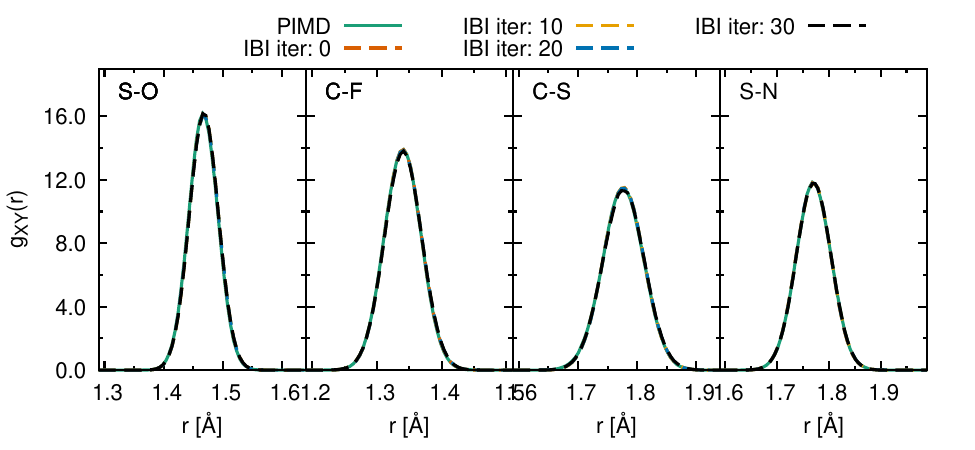} 
\caption{Convergence of f-CMD intra-molecular bond lengths of LiTSFI salt of \textbf{5m} concentration across IBI iterations compared to the exact PIMD results.}
\label{fig-litfsi-bond-5m}
\addcontentsline{toc}{subsection}{Figure \ref{fig-litfsi-bond-5m}. Convergence of f-CMD intra-molecular bond lengths of Li-TFSI salt of \textbf{5m} concentration across IBI iterations. }
\end{figure}

\clearpage
\begin{figure}[!ht] 
\centering 
\includegraphics[width=0.99\textwidth]{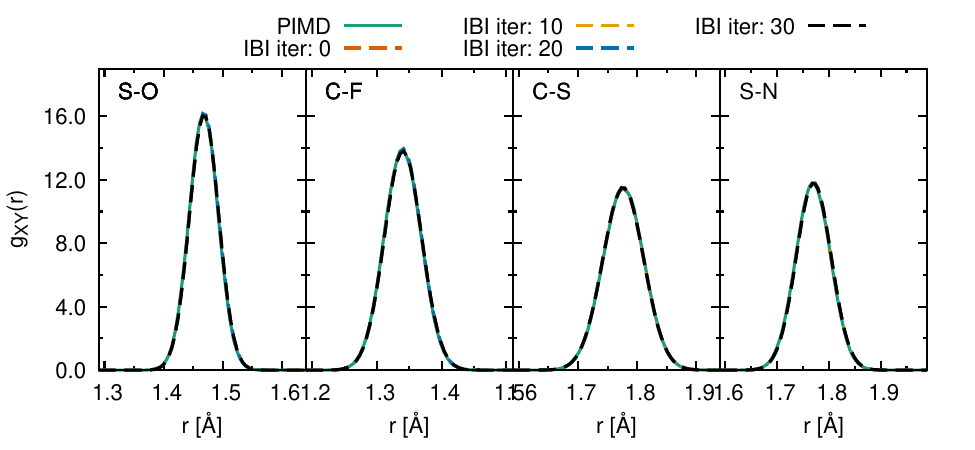} 
\caption{Convergence of f-CMD intra-molecular bond lengths of LiTSFI salt of \textbf{10m} concentration across IBI iterations compared to the exact PIMD results.}
\label{fig-litfsi-bond-10m}
\addcontentsline{toc}{subsection}{Figure \ref{fig-litfsi-bond-10m}. Convergence of f-CMD intra-molecular bond lengths of Li-TFSI salt of \textbf{10m} concentration across IBI iterations.}
\end{figure}

\clearpage
\begin{figure}[!ht] 
\centering 
\includegraphics[width=0.99\textwidth]{../figures/litfsi-bond-20m.pdf} 
\caption{Convergence of f-CMD intra-molecular bond lengths of LiTSFI salt of \textbf{20m} concentration across IBI iterations compared to the exact PIMD results.}
\label{fig-litfsi-bond-20m}
\addcontentsline{toc}{subsection}{Figure \ref{fig-litfsi-bond-20m}. Convergence of f-CMD intra-molecular bond lengths of Li-TFSI salt of \textbf{20m} concentration across IBI iterations. }
\end{figure}

\clearpage
\begin{figure}[!ht] 
\centering 
\includegraphics[width=0.99\textwidth]{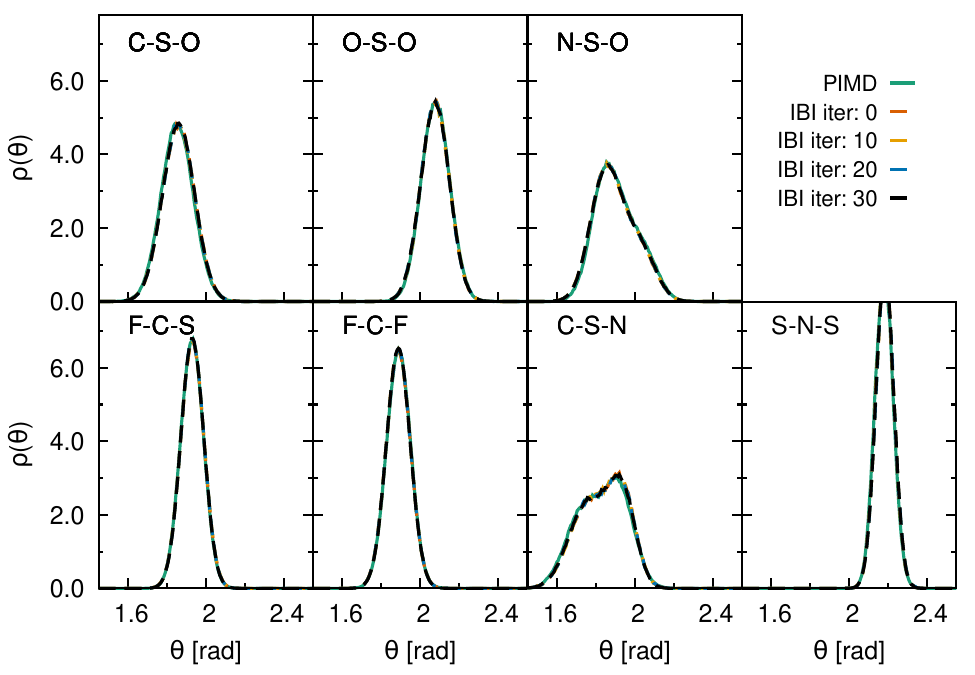} 
\caption{Convergence of f-CMD intra-molecular bond angles of LiTSFI salt of \textbf{1m} concentration across IBI iterations compared to the exact PIMD results.}
\label{fig-litfsi-angle-1m}
\addcontentsline{toc}{subsection}{Figure \ref{fig-litfsi-angle-1m}. Convergence of f-CMD intra-molecular bond angle distributions of Li-TFSI salt of \textbf{1m} concentration across IBI iterations.}
\end{figure}

\clearpage
\begin{figure}[!ht] 
\centering 
\includegraphics[width=0.99\textwidth]{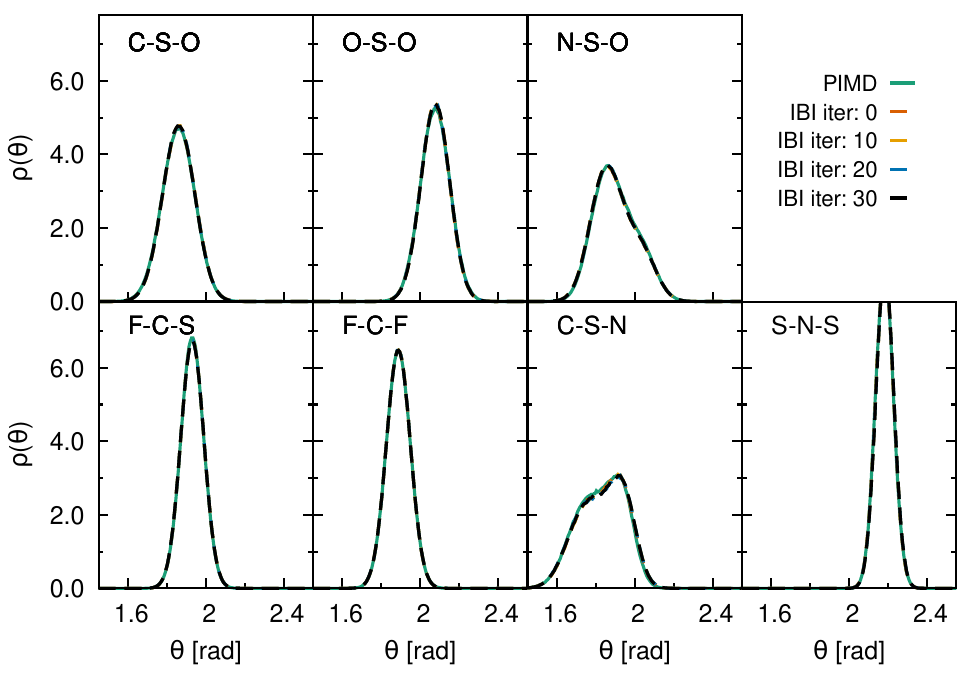} 
\caption{Convergence of f-CMD intra-molecular bond angles of LiTSFI salt of \textbf{5m} concentration across IBI iterations compared to the exact PIMD results.}
\label{fig-litfsi-angle-5m}
\addcontentsline{toc}{subsection}{Figure \ref{fig-litfsi-angle-5m}. Convergence of f-CMD intra-molecular bond angle distributions of Li-TFSI salt of \textbf{5m} concentration across IBI iterations.}
\end{figure}

\clearpage
\begin{figure}[!ht] 
\centering 
\includegraphics[width=0.99\textwidth]{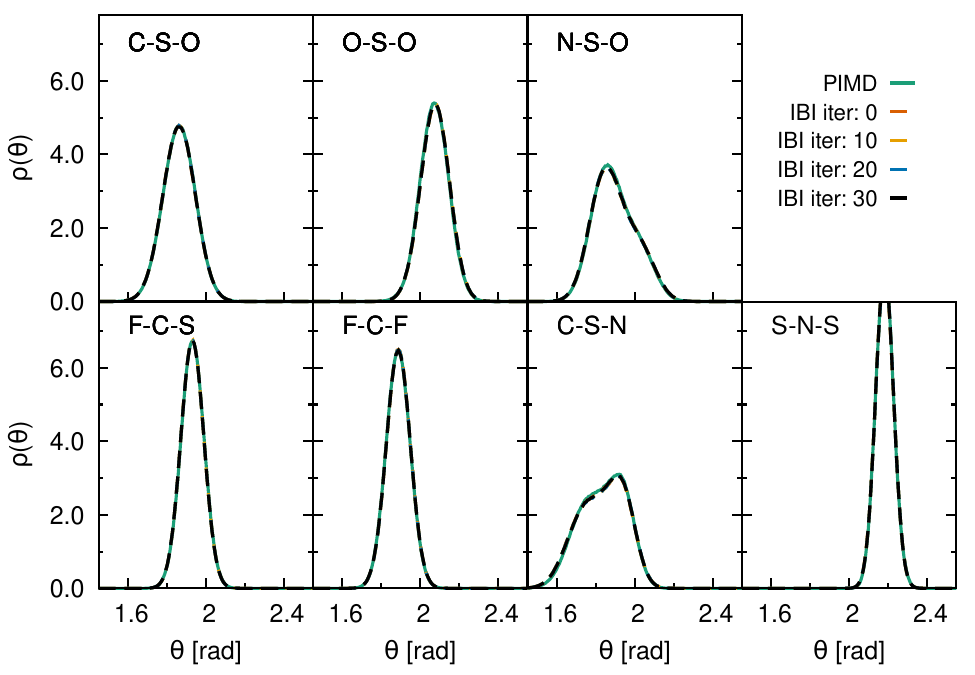} 
\caption{Convergence of f-CMD intra-molecular bond angles of LiTSFI salt of \textbf{10m} concentration across IBI iterations compared to the exact PIMD results.}
\label{fig-litfsi-angle-10m}
\addcontentsline{toc}{subsection}{Figure \ref{fig-litfsi-angle-10m}. Convergence of f-CMD intra-molecular bond angle distributions of Li-TFSI salt of \textbf{10m} concentration across IBI iterations.}
\end{figure}

\clearpage
\begin{figure}[!ht] 
\centering 
\includegraphics[width=0.99\textwidth]{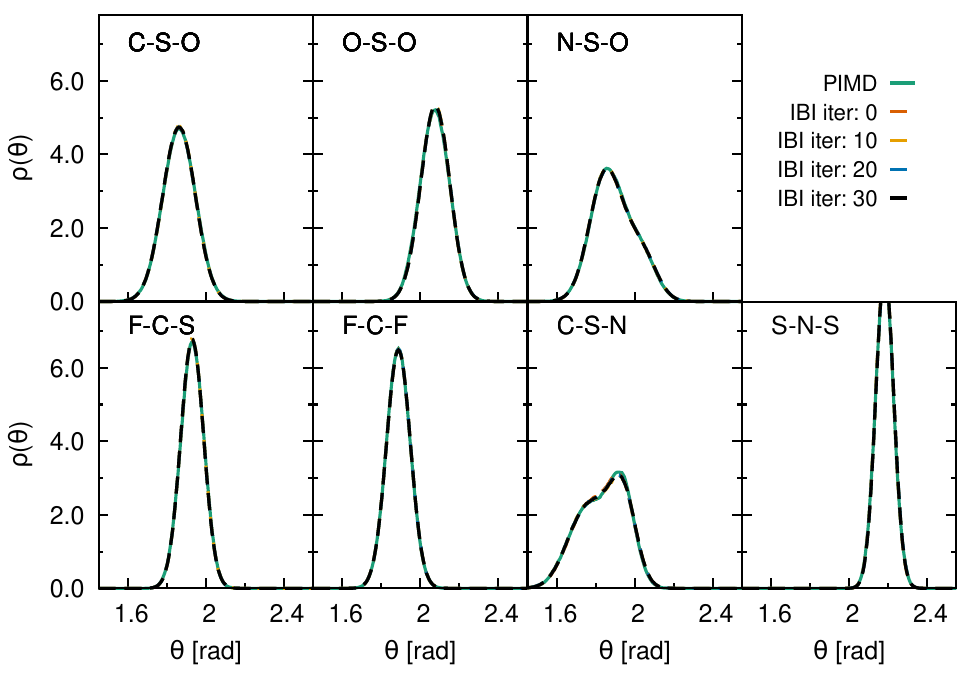} 
\caption{Convergence of f-CMD intra-molecular bond angles of LiTSFI salt of \textbf{20m} concentration across IBI iterations compared to the exact PIMD results.}
\label{fig-litfsi-angle-20m}
\addcontentsline{toc}{subsection}{Figure \ref{fig-litfsi-angle-20m}. Convergence of f-CMD intra-molecular bond angle distributions of Li-TFSI salt of \textbf{20m} concentration across IBI iterations.}
\end{figure}

\clearpage
\begin{figure}[!ht] 
\centering 
\includegraphics[width=0.99\textwidth]{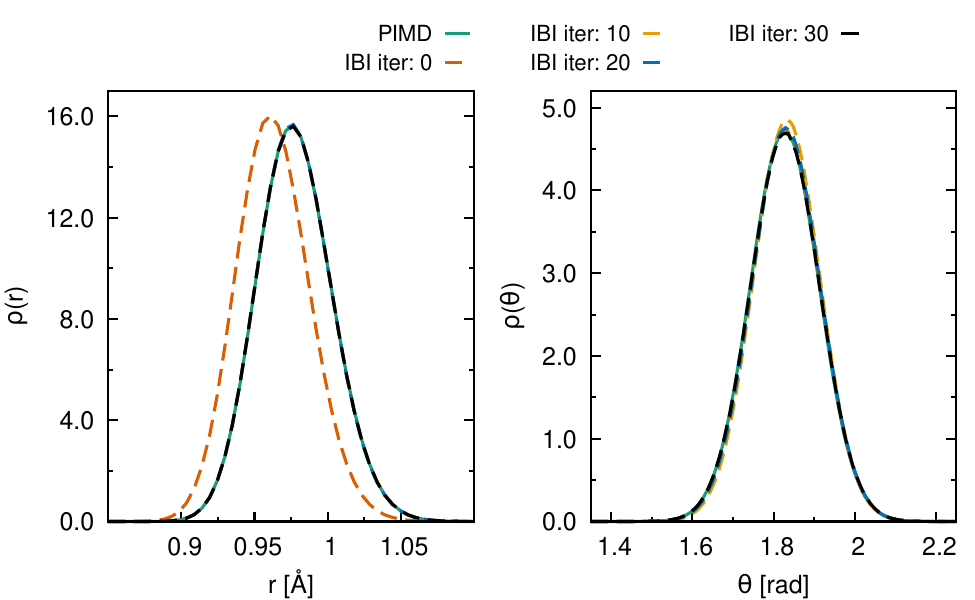} 
\caption{Convergence of f-QCMD intra-molecular distribution functions of water in \textbf{1m} aqueous Li-TFSI electrolyte.}
\label{fig-water-qc-1m}
\addcontentsline{toc}{subsection}{Figure \ref{fig-water-qc-1m}. Convergence of f-QCMD intra-molecular distribution functions of water in \textbf{1m} aqueous Li-TFSI electrolyte.}
\end{figure}

\clearpage
\begin{figure}[!ht] 
\centering 
\includegraphics[width=0.99\textwidth]{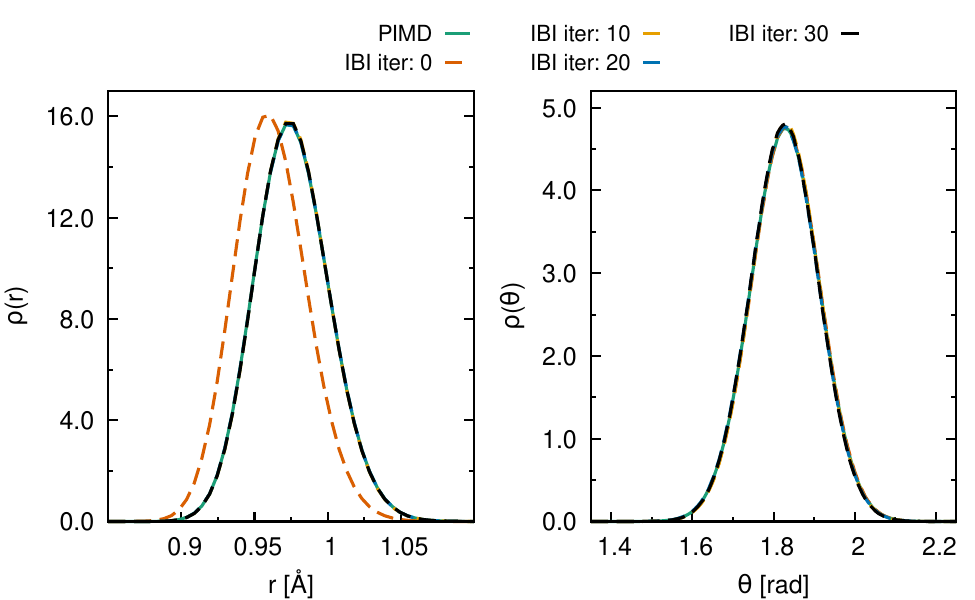} 
\caption{Convergence of f-QCMD intra-molecular distribution functions of water in \textbf{5m} aqueous Li-TFSI electrolyte.}
\label{fig-water-qc-5m}
\addcontentsline{toc}{subsection}{Figure \ref{fig-water-qc-5m}. Convergence of f-QCMD intra-molecular distribution functions of water in \textbf{5m} aqueous Li-TFSI electrolyte.}
\end{figure}

\clearpage
\begin{figure}[!ht] 
\centering 
\includegraphics[width=0.99\textwidth]{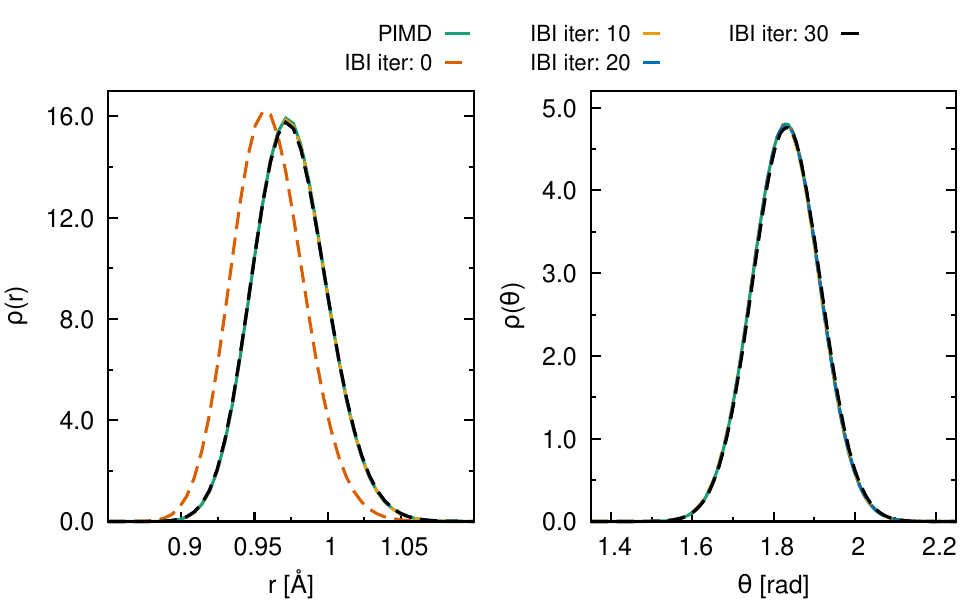} 
\caption{Convergence of f-QCMD intra-molecular distribution functions of water in \textbf{10m} aqueous Li-TFSI electrolyte.}
\label{fig-water-qc-10m}
\addcontentsline{toc}{subsection}{Figure \ref{fig-water-qc-10m}. Convergence of f-QCMD intra-molecular distribution functions of water in \textbf{10m} aqueous Li-TFSI electrolyte.}
\end{figure}

\clearpage
\begin{figure}[!ht] 
\centering 
\includegraphics[width=0.99\textwidth]{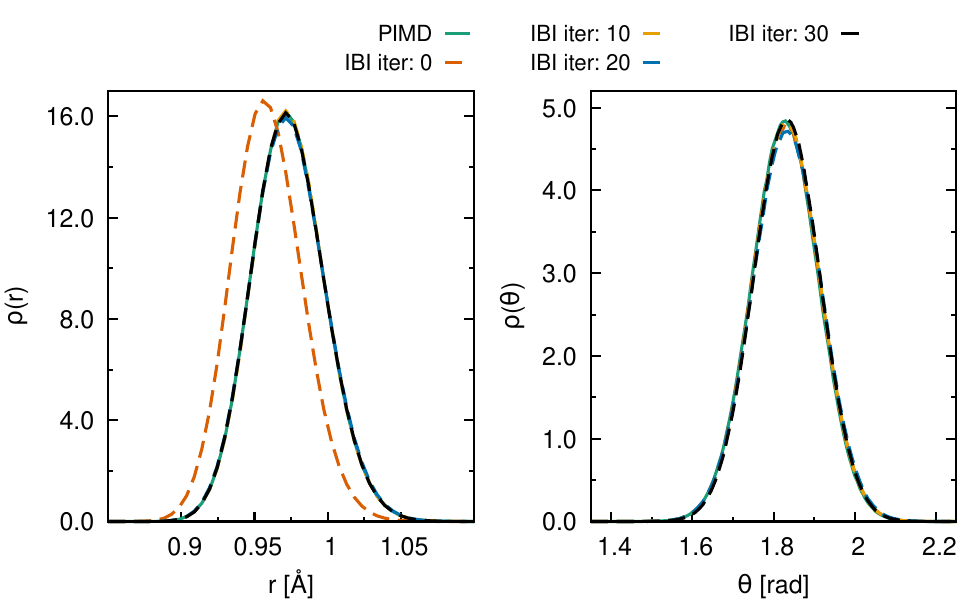} 
\caption{Convergence of f-QCMD intra-molecular distribution functions of water in \textbf{20m} aqueous Li-TFSI electrolyte.}
\label{fig-water-qc-20m}
\addcontentsline{toc}{subsection}{Figure \ref{fig-water-qc-20m}. Convergence of f-QCMD intra-molecular distribution functions of water in \textbf{20m} aqueous Li-TFSI electrolyte.}
\end{figure}

\clearpage
\begin{figure}[!ht] 
\centering 
\includegraphics[width=0.99\textwidth]{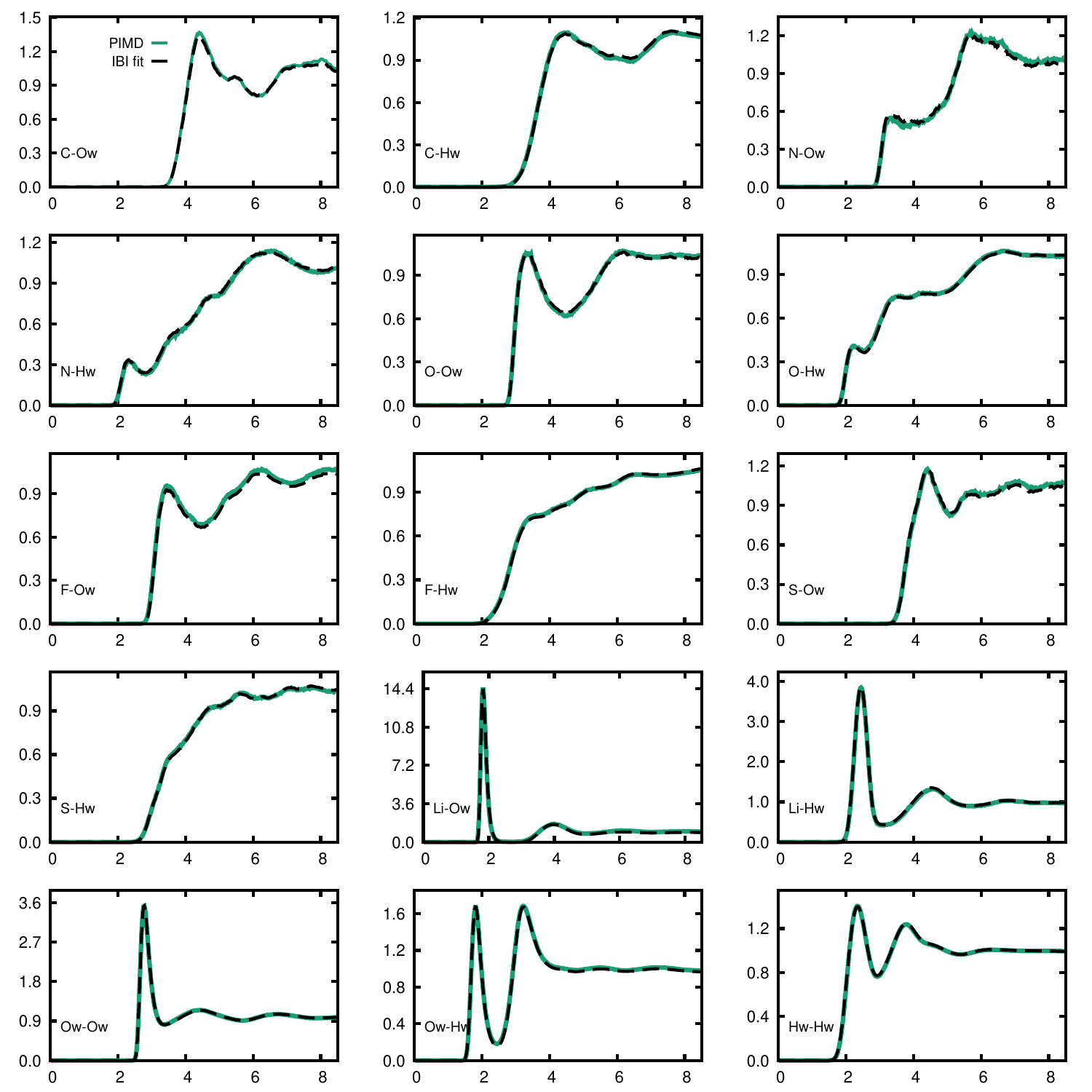} 
\caption{Convergence of f-CMD radial distribution functions for all the pairs of aqueous Li-TFSI electrolytes of \textbf{1m} solution over IBI iterations compared to the exact PIMD results.}
\label{fig-rdf-1m}
\addcontentsline{toc}{subsection}{Figure \ref{fig-rdf-1m}. Convergence of f-CMD radial distribution functions for the \textbf{1m} aqueous Li-TFSI electrolyte.}
\end{figure}

\clearpage
\begin{figure}[!ht] 
\centering 
\includegraphics[width=0.99\textwidth]{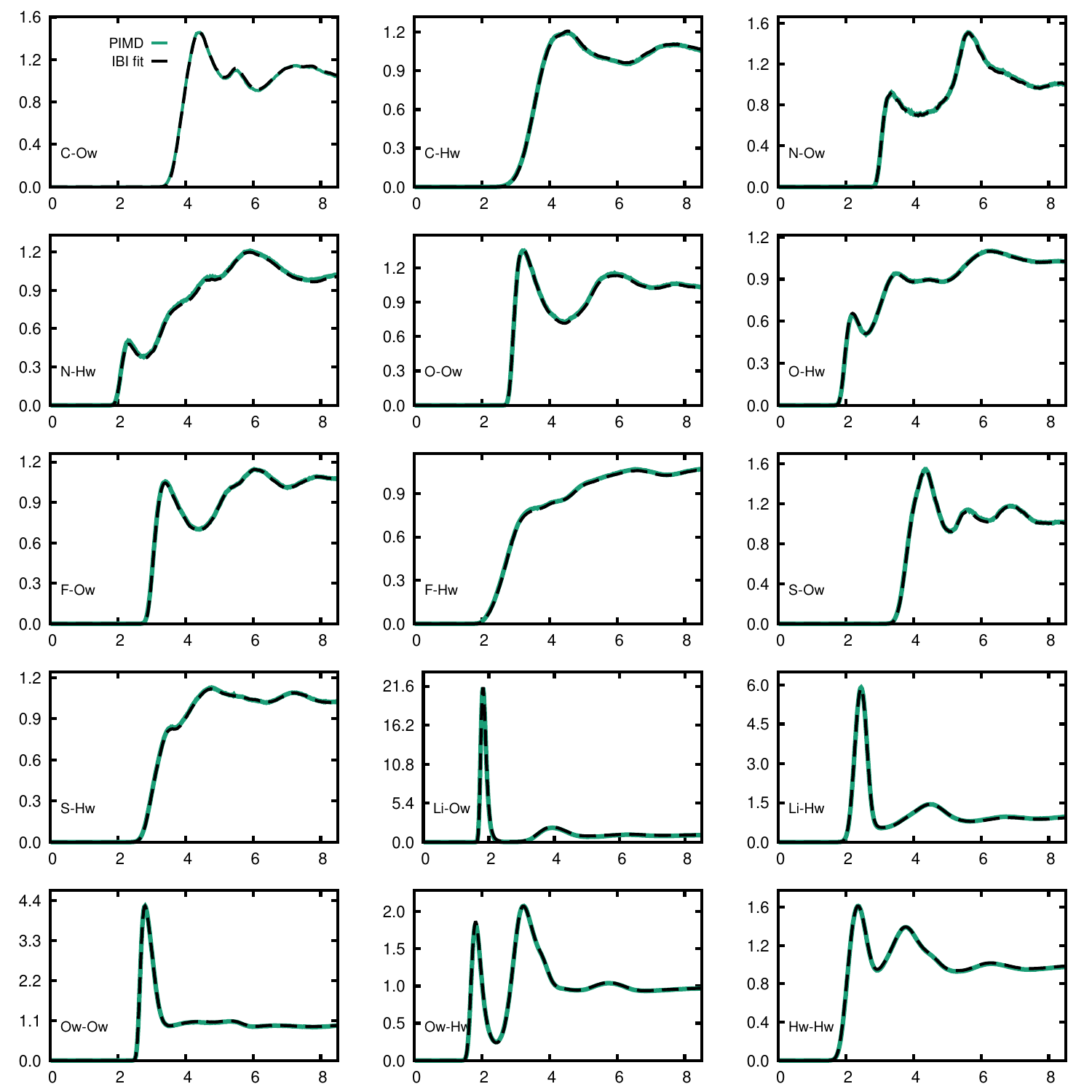} 
\caption{Convergence of f-CMD radial distribution functions for all the pairs of aqueous Li-TFSI electrolytes of \textbf{5m} solution over IBI iterations compared to the exact PIMD results.}
\label{fig-rdf-5m}
\addcontentsline{toc}{subsection}{Figure \ref{fig-rdf-5m}. Convergence of f-CMD radial distribution functions for the \textbf{5m} aqueous Li-TFSI electrolyte}
\end{figure}

\clearpage
\begin{figure}[!ht] 
\centering 
\includegraphics[width=0.99\textwidth]{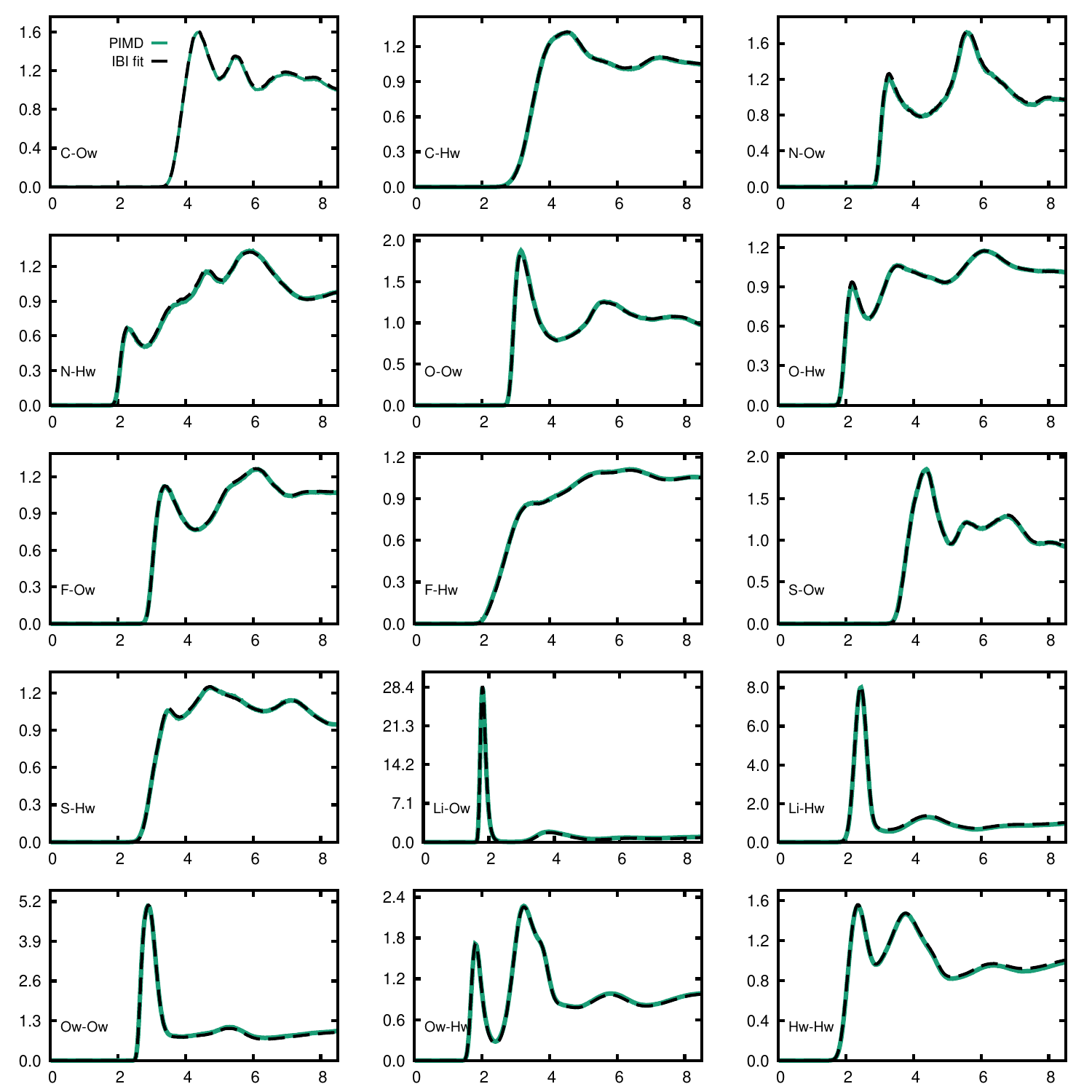} 
\caption{Convergence of f-CMD radial distribution functions for all the pairs of aqueous Li-TFSI electrolytes of \textbf{10m} solution over IBI iterations compared to the exact PIMD results.}
\label{fig-rdf-10m}
\addcontentsline{toc}{subsection}{Figure \ref{fig-rdf-10m}. Convergence of f-CMD radial distribution functions for the \textbf{10m} aqueous Li-TFSI electrolyte}
\end{figure}

\clearpage
\begin{figure}[!ht] 
\centering 
\includegraphics[width=0.99\textwidth]{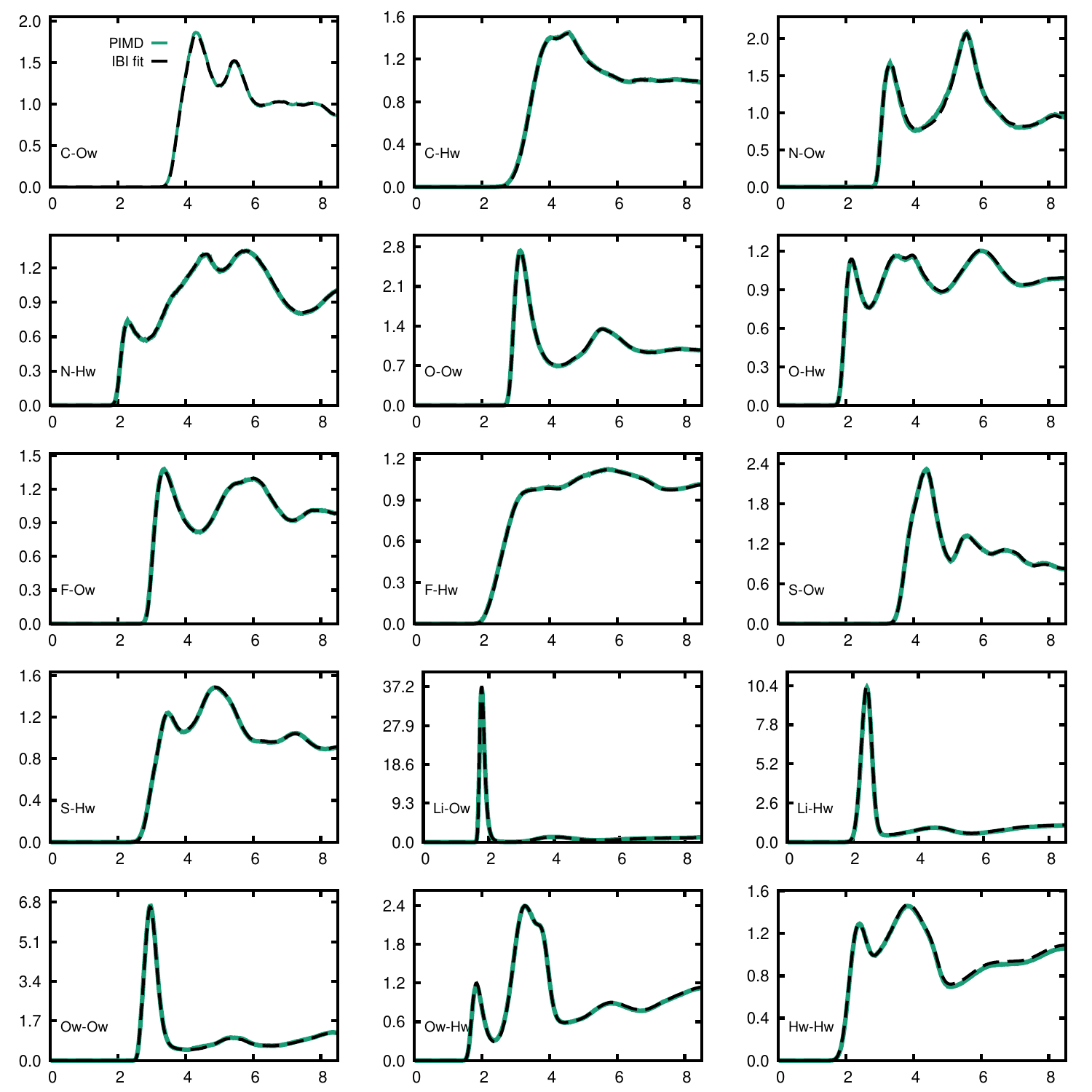} 
\caption{Convergence of f-CMD radial distribution functions for all the pairs of aqueous Li-TFSI electrolytes of \textbf{20m} solution over IBI iterations compared to the exact PIMD results.}
\label{fig-rdf-20m}
\addcontentsline{toc}{subsection}{Figure \ref{fig-rdf-20m}. Convergence of f-CMD radial distribution functions for the \textbf{20m} aqueous Li-TFSI electrolyte}
\end{figure}

\clearpage
\begin{figure}[!ht] 
\centering 
\includegraphics[width=0.7\textwidth]{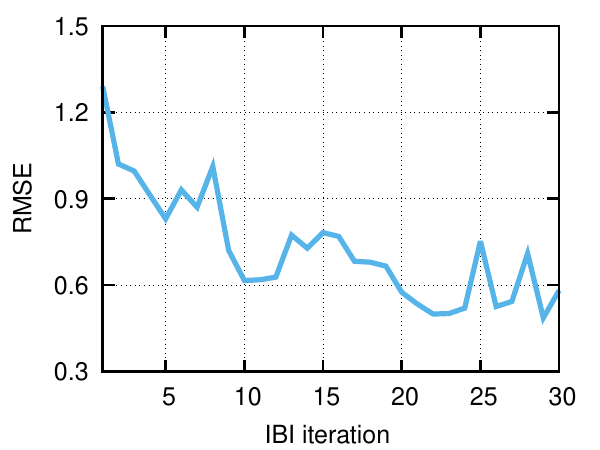} 
\caption{Convergence of the RMSE between reference and computed RDFs across IBI iterations for the \textbf{20m} case. Radial distribution functions for all 15 pairs are used to calculate the RMSE.}
\label{fig-rmse}
\addcontentsline{toc}{subsection}{Figure \ref{fig-rmse}. Convergence of RMSE over IBI iteration for the \textbf{20m} case.}
\end{figure}

\clearpage
\begin{figure}[!ht] 
\centering 
\includegraphics[width=0.99\textwidth]{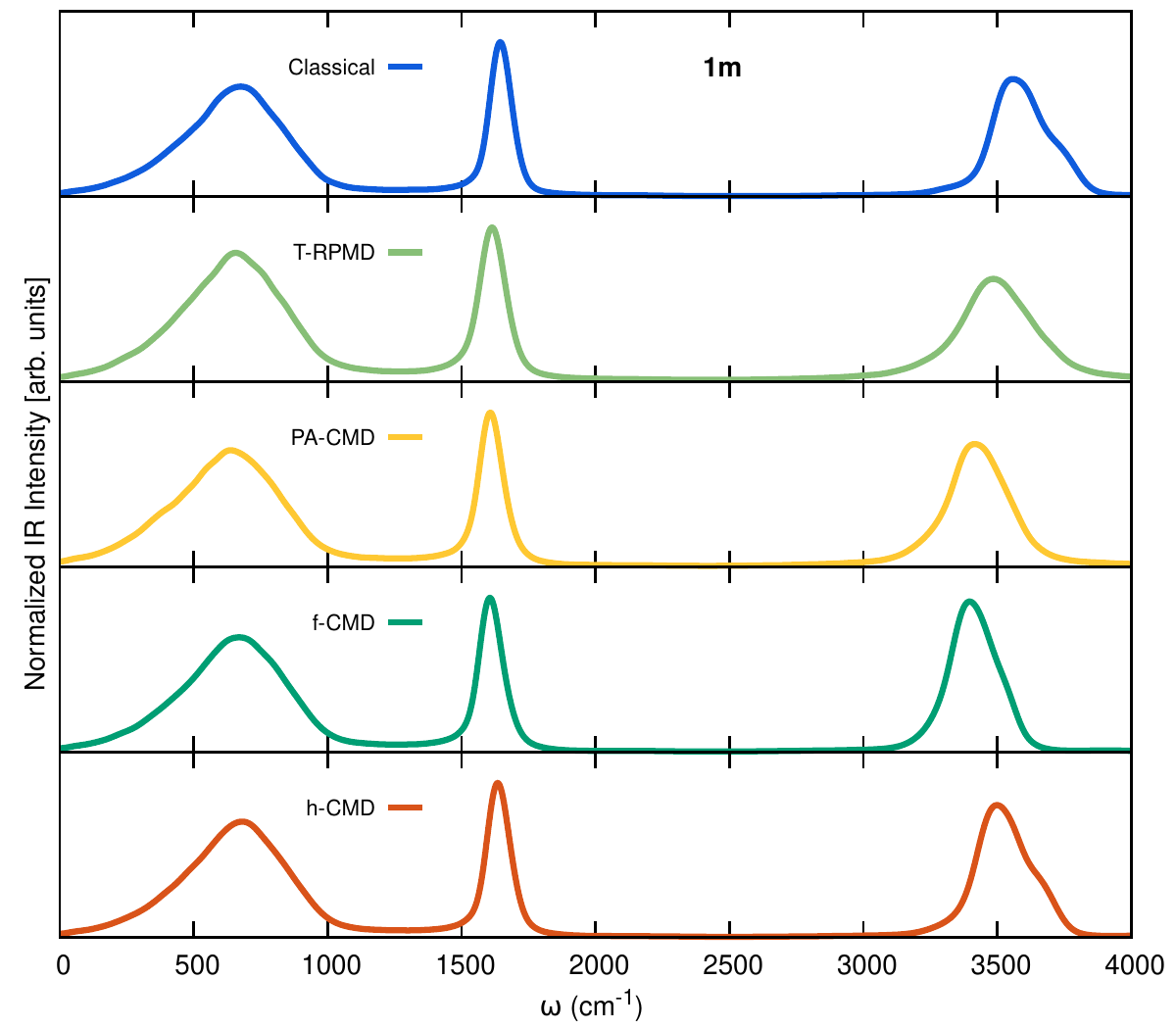} 
    \caption{Calculated full IR spectra for the \textbf{1m} aqueous Li-TFSI electrolyte solution at 298 K using different simulation methods.}
\label{fig:spec-1m}
\addcontentsline{toc}{subsection}{Figure \ref{fig:spec-1m}. Full IR spectra for \textbf{1m} aqueous Li-TFSI electrolyte.}
\end{figure}

\clearpage
\begin{figure}[!ht] 
\centering 
\includegraphics[width=0.99\textwidth]{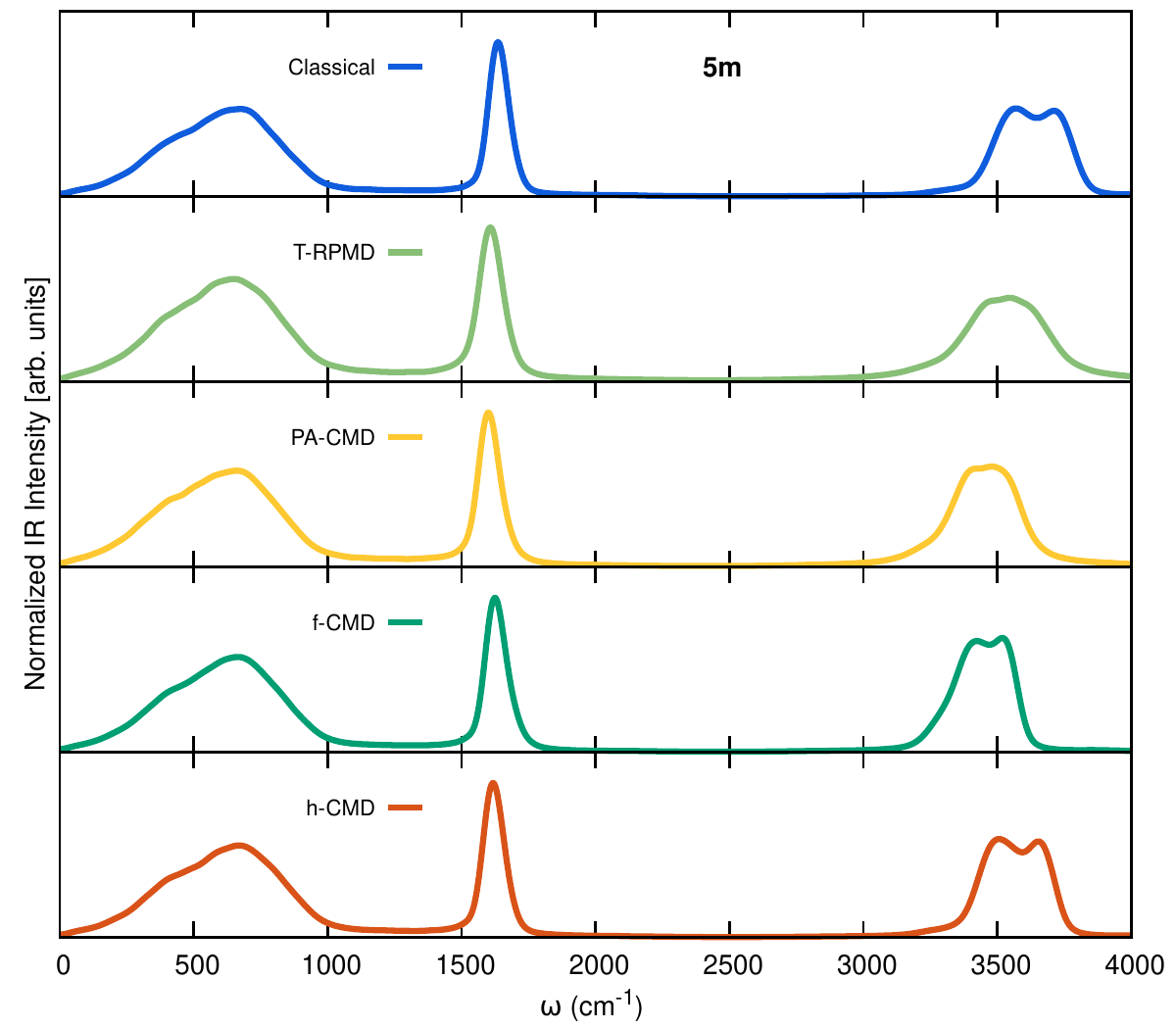} 
    \caption{Calculated full IR spectra for the \textbf{5m} aqueous Li-TFSI electrolyte solution at 298 K using different simulation methods.}
\label{fig:spec-5m}
\addcontentsline{toc}{subsection}{Figure \ref{fig:spec-5m}. Full IR spectra for \textbf{5m} aqueous Li-TFSI electrolyte.}
\end{figure}

\clearpage
\begin{figure}[!ht] 
\centering 
\includegraphics[width=0.99\textwidth]{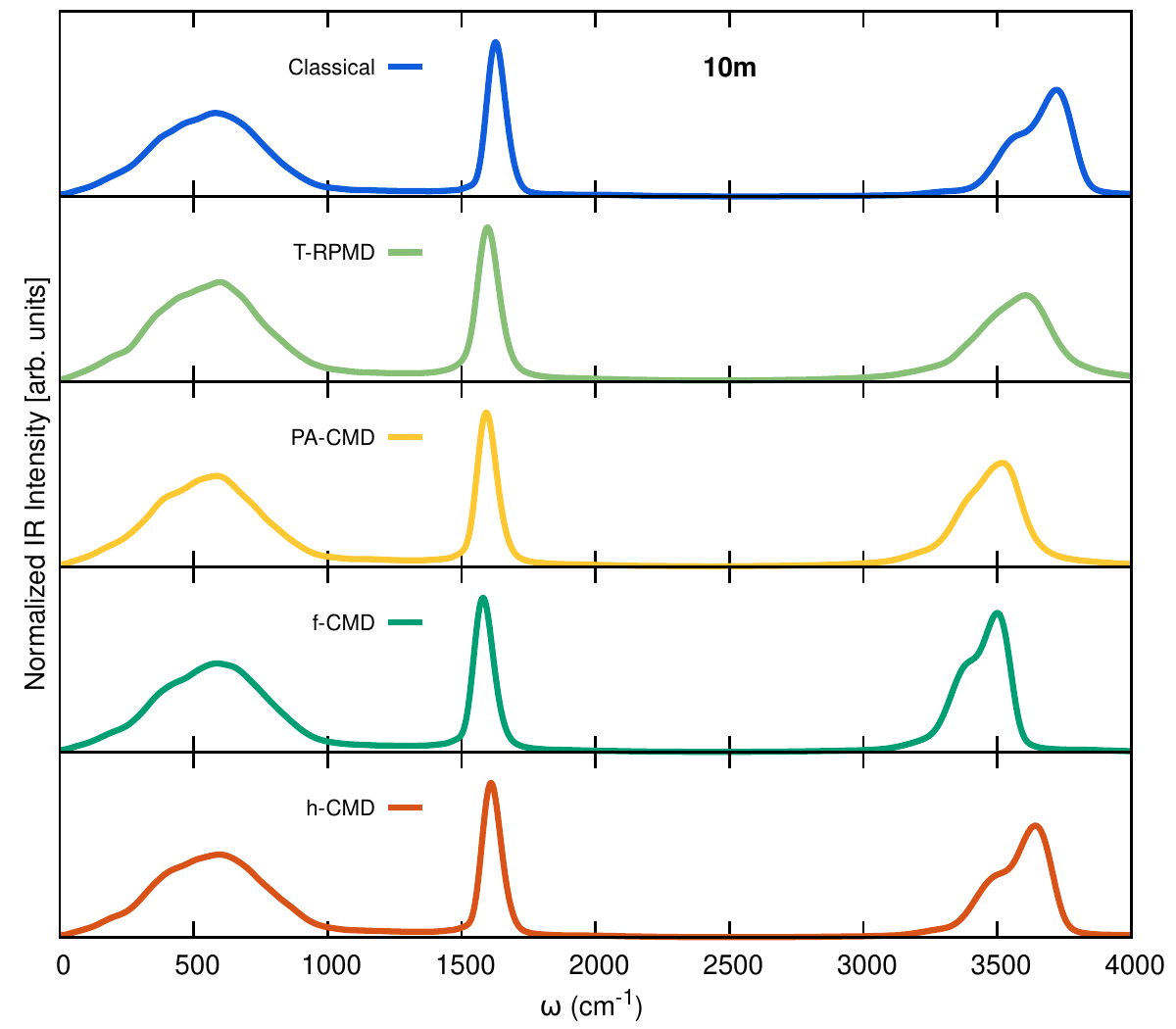} 
    \caption{Calculated full IR spectra for the \textbf{10m} aqueous Li-TFSI electrolyte solution at 298 K using different simulation methods.}
\label{fig:spec-10m}
\addcontentsline{toc}{subsection}{Figure \ref{fig:spec-10m}. Full IR spectra for \textbf{10m} aqueous Li-TFSI electrolyte.}
\end{figure}

\clearpage
\begin{figure}[!ht] 
\centering 
\includegraphics[width=0.99\textwidth]{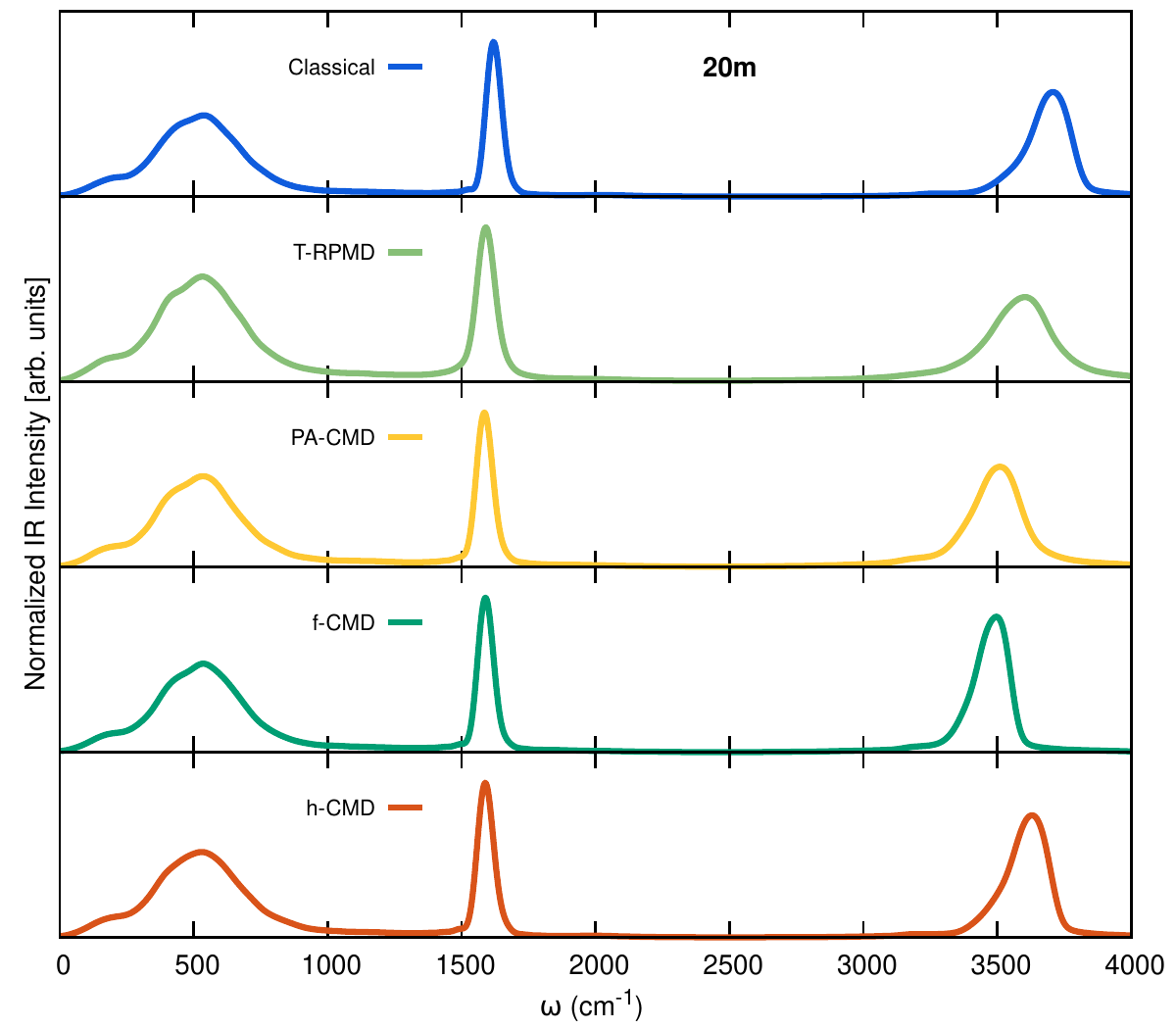} 
    \caption{Calculated full IR spectra for the \textbf{20m} aqueous Li-TFSI electrolyte solution at 298 K using different simulation methods.}
\label{fig:spec-20m}
\addcontentsline{toc}{subsection}{Figure \ref{fig:spec-20m}. Full IR spectra for \textbf{20m} aqueous Li-TFSI electrolyte.}
\end{figure}

\clearpage
\begin{figure}[!ht] 
\centering 
\includegraphics[width=0.99\linewidth]{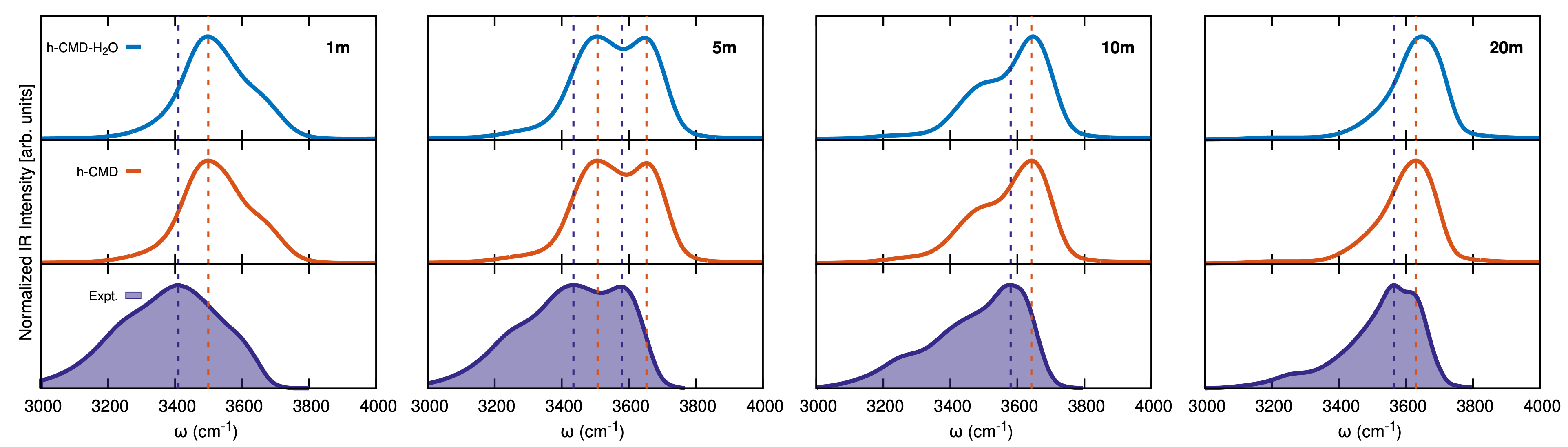}
\caption{Simulated IR spectra for different aqueous Li-TFSI electrolyte solutions, comparing the approximate h-CMD-H$_2$O method to the full h-CMD and the experimental spectra from Ref~\citenum{litfsi_spectra}. Purple and red vertical dashed lines represent experimental and h-CMD peak positions, respectively.}
\label{spec-hcmd-h2o-all}
\addcontentsline{toc}{subsection}{Figure \ref{spec-hcmd-h2o-all}. IR spectra for different aqueous Li-TFSI electrolyte solutions comparing the approximate h-CMD-H$_2$O method to the full h-CMD method and the experiment.}
\end{figure}

\clearpage
\begin{table}
\caption{Average simulation temperature of \textbf{1m} aqueous Li-TFSI salt solution with different simulation methods for all 20 NVE trajectories.}
\addcontentsline{toc}{subsection}{Table \ref{tab:1m-avg-t}. Average simulation temperature of \textbf{1m} aqueous Li-TFSI salt solution.}
\centering
\setlength{\tabcolsep}{4mm}
\scalebox{.95}{
\begin{tabular}{c c c c c c}
\midrule[1.4pt]
Trajectory & MD  & T-RPMD  & PA-CMD & f-CMD & h-CMD  \\
\midrule[1.4pt]
  1 &  298.3 &  297.9 &  298.3 &  292.1 &  304.1 \\
  2 &  300.3 &  297.9 &  298.8 &  299.1 &  296.3 \\
  3 &  300.6 &  297.9 &  298.3 &  301.5 &  295.8 \\
  4 &  296.4 &  297.9 &  298.6 &  296.8 &  293.1 \\
  5 &  301.4 &  297.8 &  298.8 &  297.2 &  289.6 \\
  6 &  299.4 &  298.0 &  298.7 &  302.8 &  293.2 \\
  7 &  298.9 &  297.9 &  298.5 &  296.2 &  293.9 \\
  8 &  294.3 &  297.9 &  298.7 &  302.8 &  297.2 \\
  9 &  301.0 &  297.9 &  298.7 &  295.5 &  295.1 \\
 10 &  300.7 &  298.0 &  298.4 &  298.1 &  290.7 \\
 11 &  295.1 &  298.0 &  298.9 &  301.2 &  299.4 \\
 12 &  301.5 &  297.9 &  298.8 &  297.0 &  290.8 \\
 13 &  296.7 &  298.0 &  298.5 &  296.6 &  293.1 \\
 14 &  299.2 &  297.9 &  298.6 &  296.3 &  299.5 \\
 15 &  298.1 &  298.0 &  298.6 &  298.7 &  295.9 \\
 16 &  299.9 &  297.9 &  298.7 &  304.3 &  297.9 \\
 17 &  296.0 &  297.9 &  298.8 &  300.6 &  300.9 \\
 18 &  297.2 &  298.0 &  298.6 &  296.7 &  298.5 \\
 19 &  295.8 &  298.0 &  298.6 &  299.9 &  298.2 \\
 20 &  298.1 &  298.0 &  298.4 &  298.4 &  303.2 \\
 \hline
 Average & 298.4 & 297.9 & 298.6 & 298.6 & 296.3 \\
\midrule[1.4pt]
\end{tabular}}     
\label{tab:1m-avg-t}
\end{table}

\clearpage
\begin{table}
\caption{Average simulation temperature of \textbf{5m} aqueous Li-TFSI salt solution with different simulation methods for all 20 NVE trajectories.}
\addcontentsline{toc}{subsection}{Table \ref{tab:5m-avg-t}. Average simulation temperature of \textbf{5m} aqueous Li-TFSI salt solution.}
\centering
\setlength{\tabcolsep}{4mm}
\scalebox{.95}{
\begin{tabular}{c c c c c c}
\midrule[1.4pt]
Trajectory & MD  & T-RPMD  & PA-CMD & f-CMD & h-CMD  \\
\midrule[1.4pt]
  1 &  299.4 &  297.9 &  298.1 &  302.2 &  292.5 \\
  2 &  298.9 &  298.0 &  298.6 &  298.1 &  297.2 \\
  3 &  298.9 &  297.9 &  298.8 &  295.4 &  292.0 \\
  4 &  296.1 &  297.9 &  298.3 &  297.2 &  298.2 \\
  5 &  297.3 &  297.9 &  298.3 &  291.7 &  298.0 \\
  6 &  299.0 &  298.0 &  298.6 &  295.6 &  293.5 \\
  7 &  294.9 &  297.9 &  298.3 &  300.7 &  299.7 \\
  8 &  296.1 &  297.8 &  298.3 &  306.1 &  299.2 \\
  9 &  299.1 &  297.8 &  298.8 &  302.2 &  302.6 \\
 10 &  303.4 &  297.9 &  298.1 &  298.4 &  301.8 \\
 11 &  293.0 &  297.9 &  298.7 &  299.6 &  298.3 \\
 12 &  295.8 &  297.9 &  298.6 &  296.6 &  298.7 \\
 13 &  299.7 &  297.8 &  298.5 &  297.1 &  297.6 \\
 14 &  296.7 &  297.8 &  298.8 &  303.0 &  299.4 \\
 15 &  296.3 &  297.7 &  298.0 &  295.0 &  301.5 \\
 16 &  305.6 &  297.9 &  298.7 &  299.8 &  300.5 \\
 17 &  298.2 &  297.9 &  298.9 &  299.0 &  296.9 \\
 18 &  296.1 &  297.9 &  298.5 &  302.6 &  292.5 \\
 19 &  295.4 &  298.0 &  298.5 &  302.0 &  296.8 \\
 20 &  299.1 &  298.0 &  298.9 &  291.0 &  294.6 \\
 \hline
 Average & 297.9 & 297.9 & 298.5 & 298.7 & 297.6 \\
\midrule[1.4pt]
\end{tabular}}     
\label{tab:5m-avg-t}
\end{table}

\clearpage
\begin{table}
\caption{Average simulation temperature of \textbf{10m} aqueous Li-TFSI salt solution with different simulation methods for all 20 NVE trajectories.}
\addcontentsline{toc}{subsection}{Table \ref{tab:10m-avg-t}. Average simulation temperature of \textbf{10m} aqueous Li-TFSI salt solution.}
\centering
\setlength{\tabcolsep}{4mm}
\scalebox{.95}{
\begin{tabular}{c c c c c c}
\midrule[1.4pt]
Trajectory & MD  & T-RPMD  & PA-CMD & f-CMD & h-CMD  \\
\midrule[1.4pt]
  1 &  301.6 &  297.9 &  298.4 &  302.6 &  302.4 \\
  2 &  300.1 &  298.0 &  298.6 &  298.8 &  297.5 \\
  3 &  299.4 &  297.9 &  298.4 &  296.2 &  297.6 \\
  4 &  301.1 &  297.8 &  298.5 &  298.4 &  295.6 \\
  5 &  293.7 &  297.9 &  298.5 &  300.4 &  293.9 \\
  6 &  297.2 &  297.9 &  298.4 &  298.3 &  294.6 \\
  7 &  300.3 &  298.0 &  298.6 &  300.3 &  298.1 \\
  8 &  299.9 &  297.9 &  298.2 &  301.8 &  294.3 \\
  9 &  295.8 &  298.0 &  298.7 &  298.0 &  293.8 \\
 10 &  298.1 &  297.9 &  298.2 &  298.8 &  290.8 \\
 11 &  298.0 &  297.9 &  298.0 &  301.7 &  298.2 \\
 12 &  294.8 &  297.9 &  298.3 &  294.5 &  296.1 \\
 13 &  305.2 &  297.7 &  297.9 &  300.9 &  299.0 \\
 14 &  301.6 &  297.8 &  298.3 &  295.9 &  300.1 \\
 15 &  297.7 &  297.8 &  298.3 &  297.1 &  300.8 \\
 16 &  296.2 &  297.9 &  298.2 &  293.6 &  294.4 \\
 17 &  300.3 &  297.9 &  298.3 &  298.3 &  291.8 \\
 18 &  293.4 &  297.9 &  298.2 &  301.7 &  297.4 \\
 19 &  292.4 &  298.0 &  298.2 &  297.2 &  299.2 \\
 20 &  292.4 &  297.9 &  298.1 &  298.8 &  296.6 \\
 \hline
 Average & 298.0 & 297.9 & 298.3 & 298.7 & 296.6 \\
\midrule[1.4pt]
\end{tabular}}     
\label{tab:10m-avg-t}
\end{table}

\clearpage
\begin{table}
\caption{Average simulation temperature of \textbf{20m} aqueous Li-TFSI salt solution with different simulation methods for all 20 NVE trajectories.}
\addcontentsline{toc}{subsection}{Table \ref{tab:20m-avg-t}. Average simulation temperature of \textbf{20m} aqueous Li-TFSI salt solution.}
\centering
\setlength{\tabcolsep}{4mm}
\scalebox{.95}{
\begin{tabular}{c c c c c c}
\midrule[1.4pt]
Trajectory & MD  & T-RPMD  & PA-CMD & f-CMD & h-CMD  \\
\midrule[1.4pt]
  1 &  293.5 &  298.1 &  297.7 &  300.9 &  293.4 \\
  2 &  295.7 &  297.9 &  298.0 &  298.7 &  297.4 \\
  3 &  302.6 &  297.9 &  298.1 &  299.6 &  304.2 \\
  4 &  296.5 &  298.0 &  298.0 &  296.6 &  296.1 \\
  5 &  295.7 &  297.9 &  298.1 &  296.8 &  301.3 \\
  6 &  300.6 &  297.9 &  298.3 &  293.7 &  294.6 \\
  7 &  301.0 &  298.0 &  298.2 &  297.0 &  297.0 \\
  8 &  292.8 &  298.1 &  298.1 &  293.2 &  300.8 \\
  9 &  296.0 &  298.0 &  298.1 &  298.7 &  301.1 \\
 10 &  292.3 &  298.1 &  297.8 &  295.4 &  299.6 \\
 11 &  303.6 &  298.0 &  298.0 &  296.3 &  302.5 \\
 12 &  294.7 &  298.1 &  298.1 &  294.9 &  300.8 \\
 13 &  297.6 &  298.0 &  298.1 &  293.3 &  300.2 \\
 14 &  297.8 &  298.0 &  297.9 &  298.2 &  300.0 \\
 15 &  296.8 &  298.0 &  297.8 &  298.4 &  308.6 \\
 16 &  299.0 &  297.9 &  298.1 &  302.4 &  294.6 \\
 17 &  299.6 &  297.9 &  297.8 &  298.8 &  297.4 \\
 18 &  303.7 &  298.0 &  297.7 &  293.6 &  298.4 \\
 19 &  296.5 &  298.0 &  297.8 &  293.4 &  300.0 \\
 20 &  294.9 &  298.1 &  298.2 &  298.7 &  298.5 \\
 \hline
 Average & 297.5 & 298.0 & 298.0 & 296.3 & 299.2 \\
\midrule[1.4pt]
\end{tabular}}     
\label{tab:20m-avg-t}
\end{table}

\clearpage
\bibliography{articles.bib}
\bibliographystyle{achemso}